\documentclass[fleqn,usenatbib]{mnras}
\usepackage{newtxtext,newtxmath}
\usepackage[T1]{fontenc}
\usepackage{ae,aecompl}
\usepackage{graphicx}	
\usepackage{amsmath}	
\usepackage{amssymb}	
\usepackage{booktabs}	
\usepackage{url}

\title[Kepler Pulsating Binaries VI]{Finding binaries from phase modulation of pulsating stars with {\it Kepler}: VI. Orbits for 10 new binaries with mischaracterised primaries}

\author[Simon J. Murphy et al.]{
Simon J. Murphy,$^{1,2}$\thanks{E-mail: simon.murphy@sydney.edu.au (SJM)}
Nicholas H. Barbara,$^{1,2}$
Daniel Hey,$^{1,2}$\and
Timothy R. Bedding,$^{1,2}$ and
Ben D. Fulcher$^{1}$
\\
$^{1}$ Sydney Institute for Astronomy (SIfA), School of Physics, University of Sydney, NSW 2006, Australia\\
$^{2}$ Stellar Astrophysics Centre, Department of Physics and Astronomy, Aarhus University, 8000 Aarhus C, Denmark\\
}

\date{Accepted XXX. Received YYY; in original form ZZZ}
\pubyear{2015}
\begin{document}
\label{firstpage}
\pagerange{\pageref{firstpage}--\pageref{lastpage}}
\maketitle

\begin{abstract}
Measuring phase modulation in pulsating stars has proved to be a highly successful way of finding binary systems. The class of pulsating main-sequence A and F variables known as $\delta$\,Scuti stars are particularly good targets for this, and the \textit{Kepler} sample of these has been almost fully exploited. However, some \textit{Kepler} $\delta$\,Scuti stars have incorrect temperatures in stellar properties catalogues, and were missed in previous analyses. We used an automated pulsation classification algorithm to find 93 new $\delta$\,Scuti pulsators among tens of thousands of F-type stars, which we then searched for phase modulation attributable to binarity. We discovered 10 new binary systems and calculated their orbital parameters, which we compared with those of binaries previously discovered in the same way.  The results suggest that some of the new companions may be white dwarfs.
\end{abstract}




\section{Introduction}

The utility of \textit{Kepler} data has expanded far beyond the primary purpose of exoplanet transit detection \citep{boruckietal2011} and the asteroseismology program \citep{gillilandetal2010a}. An unexpectedly successful use of the data has been to infer companions to stars via their influence on the stellar pulsations \citep{shibahashi&kurtz2012,murphyetal2014}. In this endeavour, the pulsations are used as clocks whose ticks arrive early or late according to the position of the star in its orbit, similar to O$-$C analyses. The time derivative of the time delays gives the radial velocity curve, and analysis of these time delays allows the orbit to be determined. Faster clocks (shorter-period pulsations) and more stable clocks allow the orbit to be determined more precisely. So far, the class of intermediate-mass stars on the main sequence known as $\delta$\,Scuti stars has proved the most valuable for pulsation-timing studies \citep{comptonetal2016,murphy2018}.

The largest sample of pulsation-timing binaries are the 341 systems with $\delta$\,Sct primaries presented in the previous (fifth) paper of this series (\citealt{murphyetal2018}; \href{https://ui.adsabs.harvard.edu/abs/2018MNRAS.474.4322M/abstract}{Paper\,V}, hereafter). That study used $\sim$12\,600 stars in or near the $\delta$\,Sct instability strip, spanning the $T_{\rm eff}$ range 6600--10\,000\,K, of which 2200 were pulsators. This relatively small number of pulsators results from the fact that, even in the middle of the instability strip, no more than 70\% of stars actually pulsate \citep{murphyetal2019}. This pulsator fraction drops precipitously at the red edge of the instability strip, while the number of stars increases greatly due to the stellar initial mass function. In other words, expanding the search for $\delta$\,Sct stars to cooler temperatures is an exercise in diminishing returns.


A small number of \textit{Kepler} $\delta$\,Sct stars appear to lie outside of the instability strip. Generally, these have incorrect temperatures in the Kepler Input Catalogue (KIC; \citealt{brownetal2011} and its subsequent revisions by \citealt{huberetal2014,mathuretal2017,bergeretal2018}). Having incorrect KIC temperatures is even more likely for binary stars, since we observe the blend of light of the two stars. The photometric colours used to determine the effective temperatures will not be representative of typical $\delta$\,Sct stars, but will instead be biased by the companions, which are typically less-massive main-sequence stars (\href{https://ui.adsabs.harvard.edu/abs/2018MNRAS.474.4322M/abstract}{Paper\,V}) with lower temperatures and redder colours. Finding these $\delta$\,Sct stars with mischaracterised temperatures by searching for pulsations in stellar light curves is a time-consuming task that is amenable to automation. In this paper, we use machine learning on 56\,860 stars with KIC temperatures cooler than those studied in \href{https://ui.adsabs.harvard.edu/abs/2018MNRAS.474.4322M/abstract}{Paper\,V} (i.e., cooler than 6600\,K) to find 93 new $\delta$\,Sct stars and search for binaries among them.

We also investigate 362 stars hotter than the upper limit of 10\,000\,K used in \href{https://ui.adsabs.harvard.edu/abs/2018MNRAS.474.4322M/abstract}{Paper\,V}. This sample is comprised of stars from other classes useful for pulsation timing, such as $\beta$\,Cephei stars (B stars pulsating in p\:modes), subdwarf B stars (pulsating in p and/or g\:modes), white dwarfs (pulsating in g\:modes), but also the less useful Slowly Pulsating B stars (g\:modes) whose long pulsation periods are less precise clocks and a large number (295) of non-pulsators or stars with amplitudes too low to be useful for pulsation timing. It is also possible that $\delta$\,Sct stars exist as the secondaries in these systems, and that their pulsations will be detectable in the \textit{Kepler} light curves, despite the significant dilution.

We describe our target selection procedure and pulsation timing method in more detail in Sect.\,\ref{sec:method}. The results are shown in Sect.\,\ref{sec:results}.

\section{Methodology}
\label{sec:method}

\subsection{Sample definition}

Our goal was to ensure that all pulsators with coherent (not stochastic) oscillations had been analysed with the phase modulation method to search for binarity. We defined our sample to have $T_{\rm eff} > 6000$\,K in the \citet{mathuretal2017} Stellar Properties Catalogue. This extends the sample in the \href{https://ui.adsabs.harvard.edu/abs/2018MNRAS.474.4322M/abstract}{Paper\,V} catalogue to stars between 6000 and 6600\,K (the ``cool stars'' subsample) and to stars hotter than 10\,000\,K (the ``hot stars'' subsample).
We excluded stars already analysed in \href{https://ui.adsabs.harvard.edu/abs/2018MNRAS.474.4322M/abstract}{Paper\,V}.

\subsection{Classifying cool stars by variability}
For the 56\,860 stars with temperatures between 6000\,K and 6600\,K, we used a machine-learning algorithm to isolate potential $\delta$\,Sct candidates for manual (visual) inspection.
Our approach leverages existing work (Barbara et al., \textit{in prep.}), which constructed a feature space and classification algorithm for distinguishing seven classes of light curves: six common classes of variable stars---detached eclipsing binaries, contact binaries, $\delta$\,Sct stars, $\gamma$\,Dor stars, rotational variables, and RR Lyrae stars---as well as an additional class containing non-variable stars.
The training set contained 1341 Quarter 9 (Q9) \textit{Kepler} light curves of stars with $T_{\rm eff}$ between 6500 and 10\,000\,K. The light curves were obtained from KASOC\footnote{\url{http://kasoc.phys.au.dk/}} and were processed with the multi-scale MAP pipeline \citep{stumpeetal2014}.
Here we used the feature space and classification algorithm, specifically trained to accurately distinguish \textit{Kepler} light curves, to identify new $\delta$\,Sct candidates.
Detailed methods are described in Barbara et al. ({\it in prep.}) and are summarised below.

First, in identifying appropriate time-series features to represent Q9 light curve time series, we used a forward-feature selection algorithm \citep{fulcher&jones2014} on a library of over 7000 time-series features from across the scientific literature, implemented in the \textit{hctsa} package \citep{fulcheretal2013,fulcher&jones2017}.
The seven-class classification performance of each feature (or set of features) was assessed according to its balanced classification accuracy using a linear Support Vector Machine (SVM), using 10-fold cross-validation.
This procedure yielded a set of five time-series features that are well-suited to distinguishing the seven classes of light curves.
Briefly, these features measure:
\begin{enumerate}
\item non-linear autocorrelation (feature name \verb|AC_nl_012| in \textit{hctsa});
\item the probability of two magnitude increases in two successive time points (\verb|SB_MotifTwo_diff_uu|);
\item relative power contained in peaks of the Fourier transform (\verb|SP_Summaries_fft_peakPower_prom2|);
\item autocorrelation at time-lag 8 (\verb|AC_8|); and
\item entropy of the distribution of values in the time-series (\verb|EN_DistributionEntropy_hist_fd_0|).
\end{enumerate}
Representing each star in the training set in this five-dimensional feature space, we trained a Gaussian Mixture Model (GMM) to accurately classify light curves between the seven classes.

To identify new $\delta$\,Sct candidates, we converted \textit{Kepler} light curves from Q9 for the much larger number of stars in the wider $T_\mathrm{eff}$ range of 6000\,K--10\,000\,K into five-dimensional feature vectors.
We then used the trained GMM to calculate posterior probabilities for each of the seven classes and focused on the 56\,860 stars in the ``cool stars'' subsample. We generated an automated short list of 1203 candidates whose most likely variable class was $\delta$\,Sct, which included 960 $\delta$\,Sct stars already known from \citet{murphyetal2019}. By manually inspecting all stars on this list, we identified 93 new $\delta$\,Sct stars. The false-positive rate during short-listing was $\sim$9\%, with the primary sources of false positives being, in order of occurrence: harmonics of g\:mode pulsations, \textit{Kepler} data artefacts, apparent non-pulsators with noise spikes, and very low amplitude pulsators of borderline significance. Stars in the latter category could be considered $\delta$\,Sct stars, depending on the chosen significance threshold (see \citealt{murphyetal2019} for a discussion), but the amplitudes would be too low to detect any phase modulation.

Since the new $\delta$\,Sct stars have temperatures much cooler than the red edge of the $\delta$\,Sct instability strip \citep{dupretetal2005b,murphyetal2019}, their temperatures must be mischaracterised. Of the 93 $\delta$\,Sct stars, 91 had pulsation amplitudes high enough for a phase modulation analysis.

\subsection{Classifying hot stars by variability}
The hot stars subsample contained only 362 stars. The machine learning classifier was not applied to these stars because of the greater variety of variable stars and the small number of stars available to use as a training set. These stars were therefore classified manually into pulsators and non-pulsators using the full four-year \textit{Kepler} light curves. Of the 362 stars, only 51 stars had pulsations suitable for measuring phase modulation (after excluding 7 eclipsing binaries).

\subsection{Detecting and modelling binary orbits}
To search for binary stars among the 142 newly identified pulsators, we followed the method used in \href{https://ui.adsabs.harvard.edu/abs/2018MNRAS.474.4322M/abstract}{Paper\,V}, which can be summarised as follows. We calculated the Fourier transform of the four-year \textit{Kepler} light curves and used the frequencies of the strongest Fourier peaks as `clocks'. We used a non-linear least-squares algorithm to find the best-fitting frequencies, amplitudes and phases for these peaks. The resulting pulsation frequencies for each star are listed in Table\:\ref{tab:freqs} for each of the 10 systems that were ultimately found to be binaries.

\begin{figure*}
\begin{center}
\includegraphics[width=0.48\textwidth]{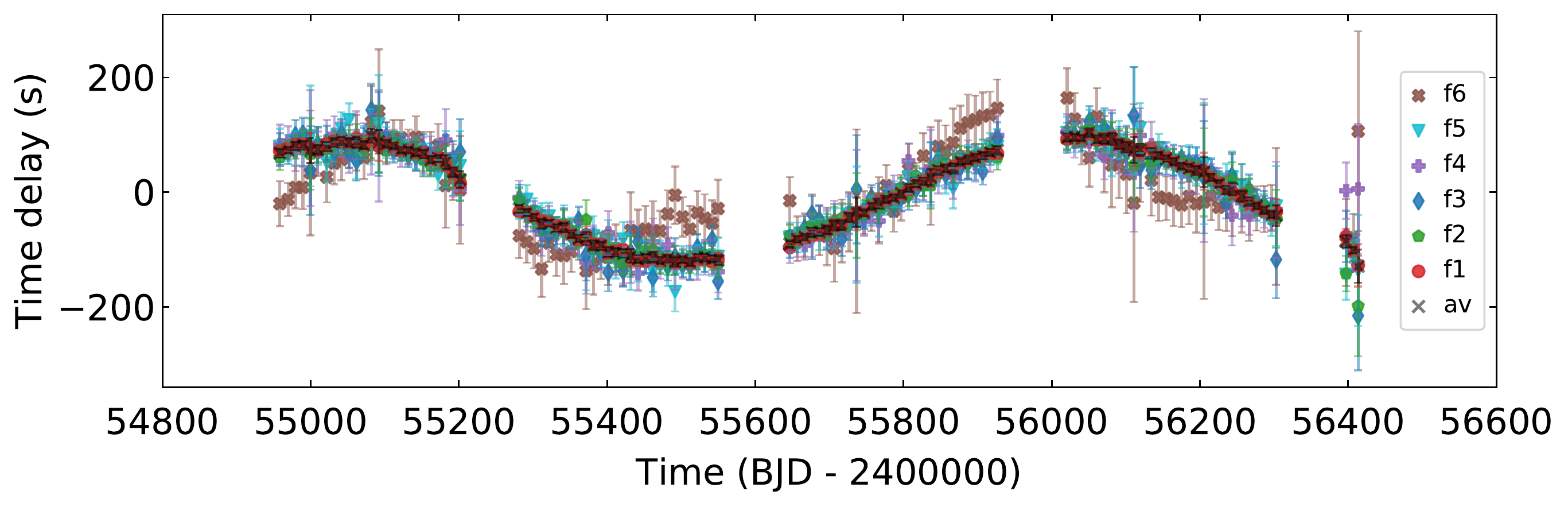}
\includegraphics[width=0.48\textwidth]{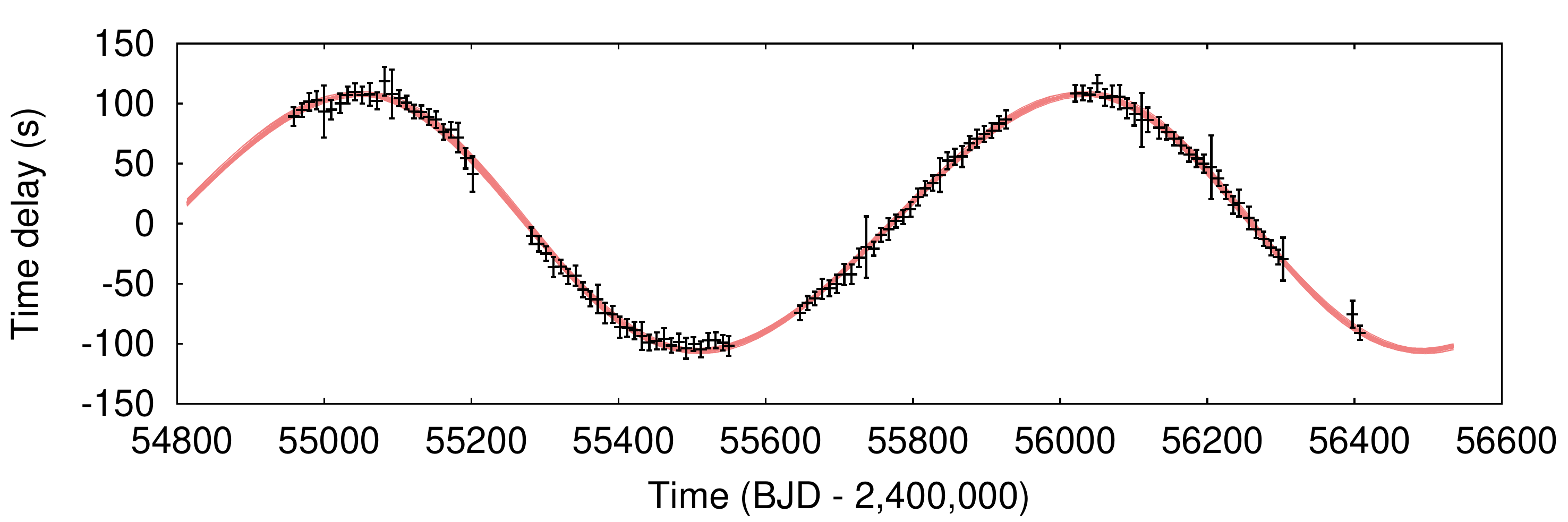}\\
\includegraphics[width=0.48\textwidth]{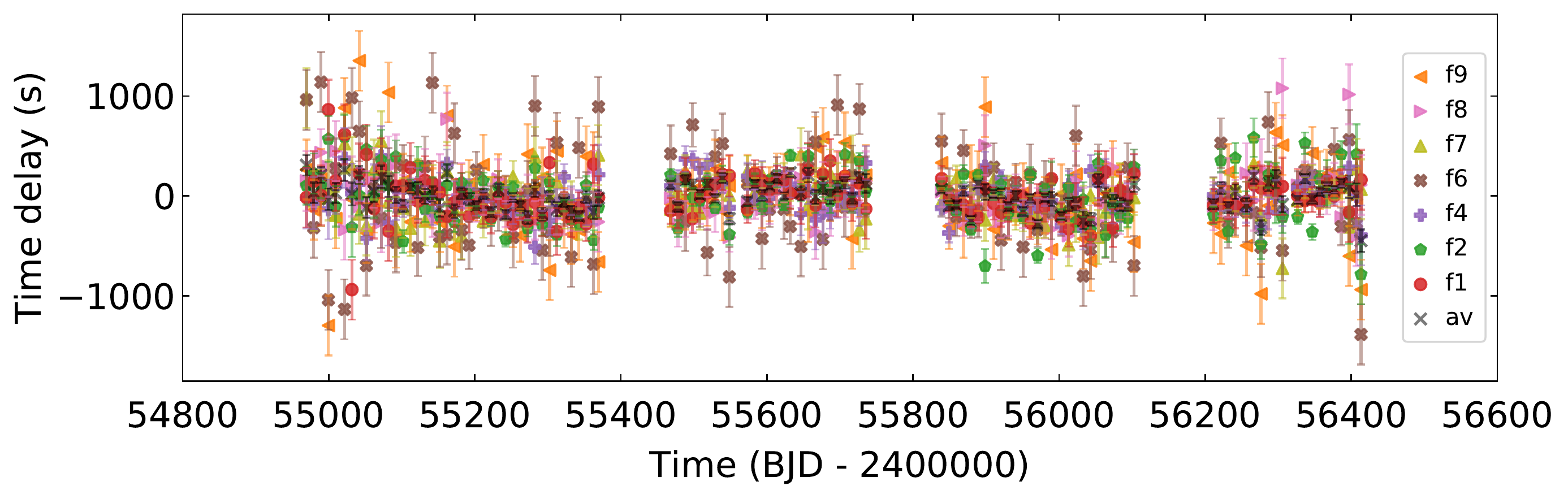}
\includegraphics[width=0.48\textwidth]{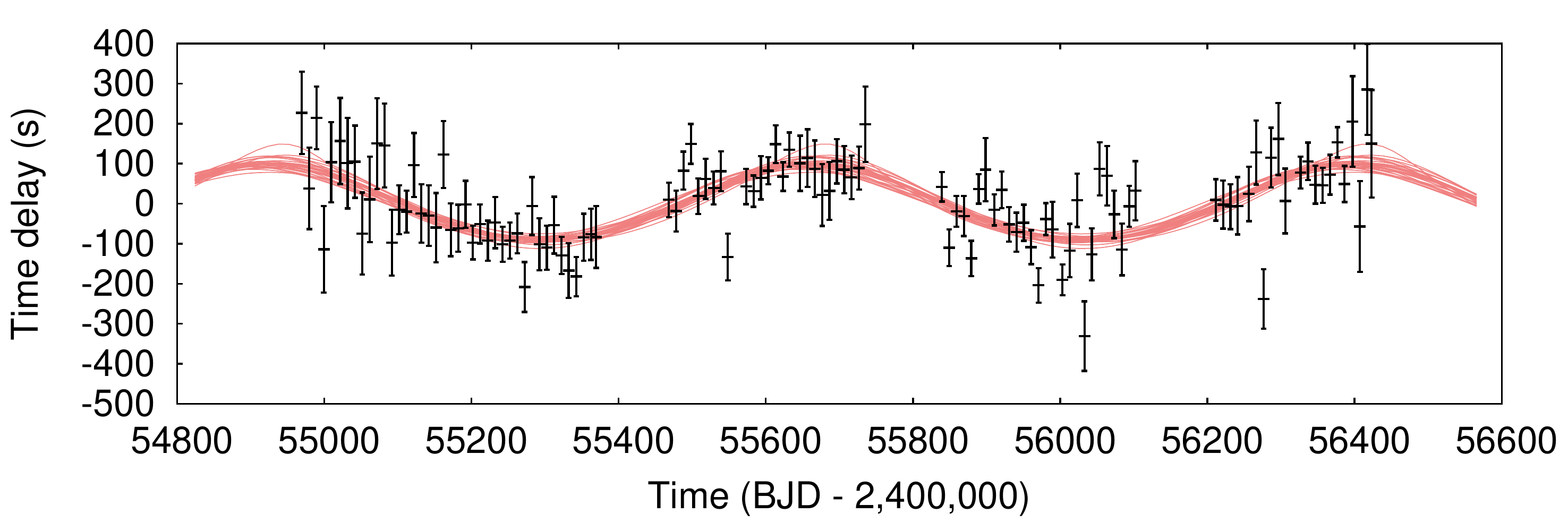}\\
\includegraphics[width=0.48\textwidth]{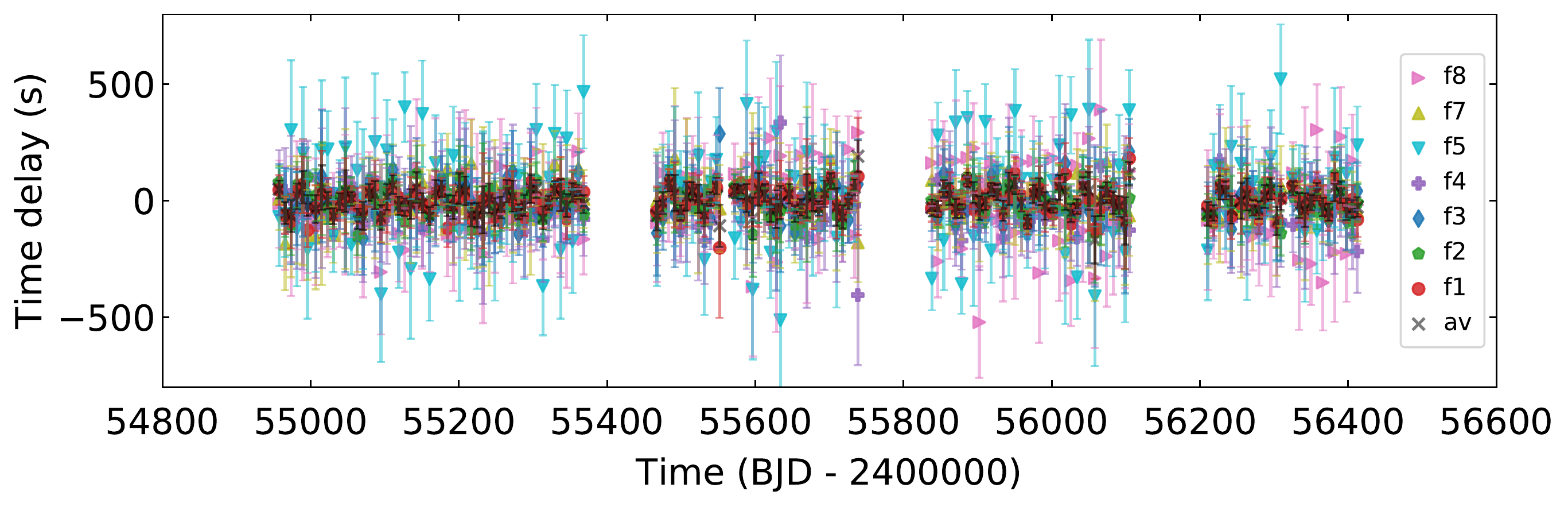}
\includegraphics[width=0.48\textwidth]{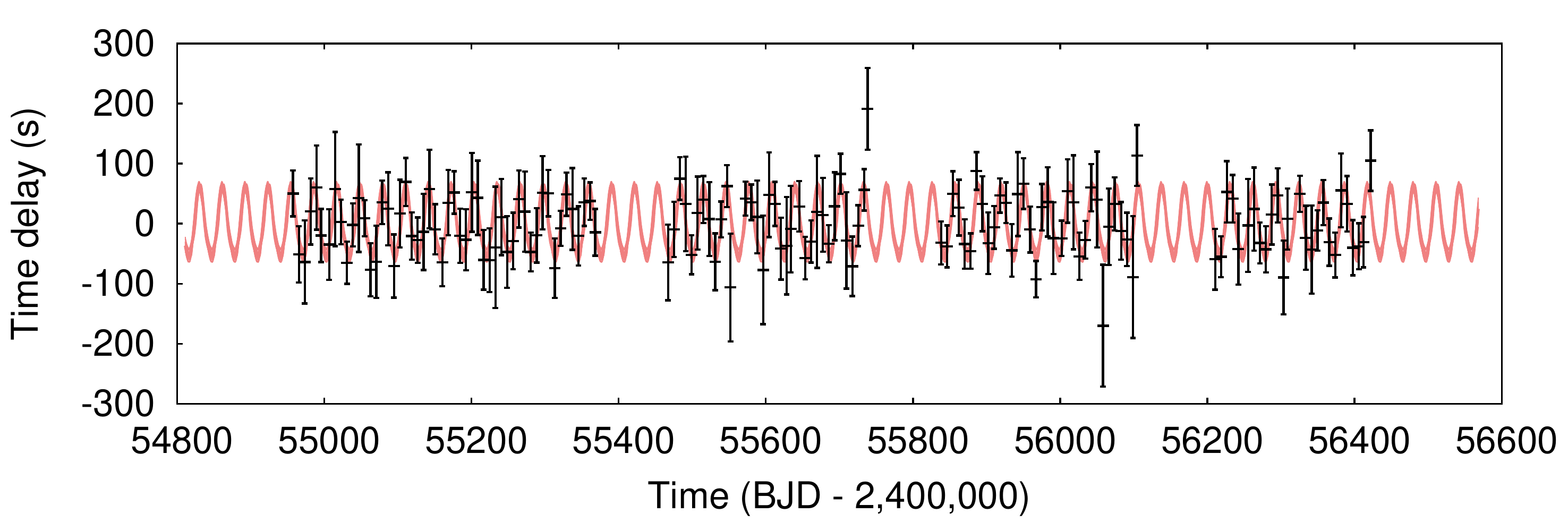}\\
\includegraphics[width=0.48\textwidth]{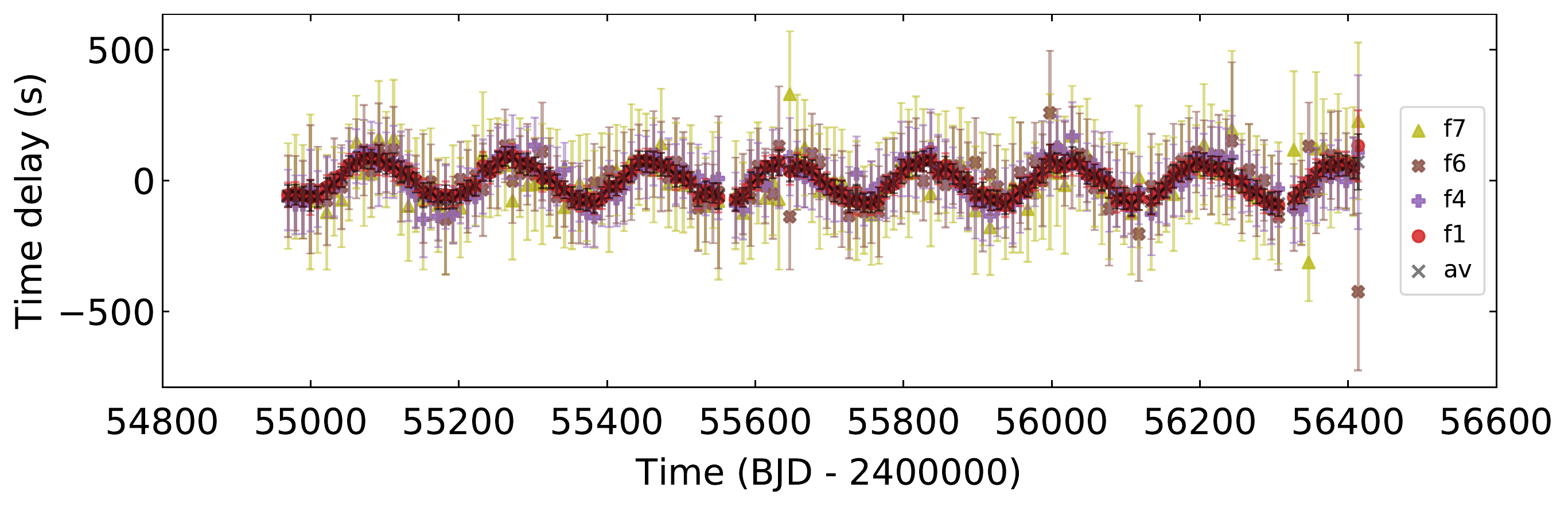}
\includegraphics[width=0.48\textwidth]{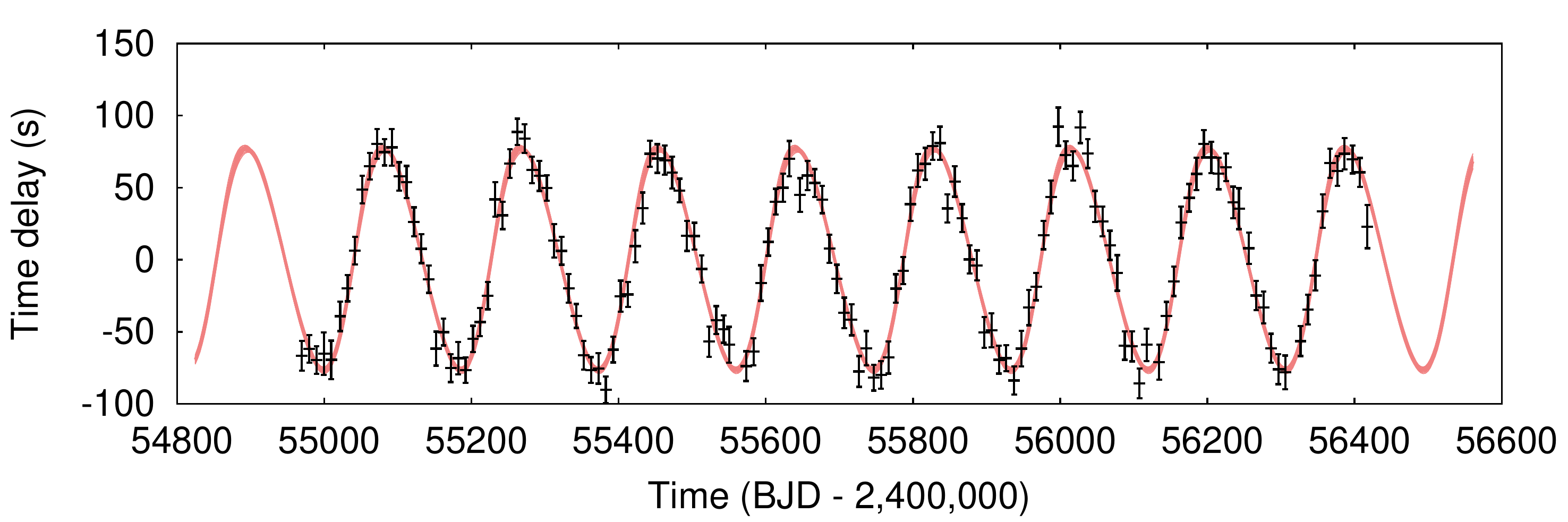}\\
\includegraphics[width=0.48\textwidth]{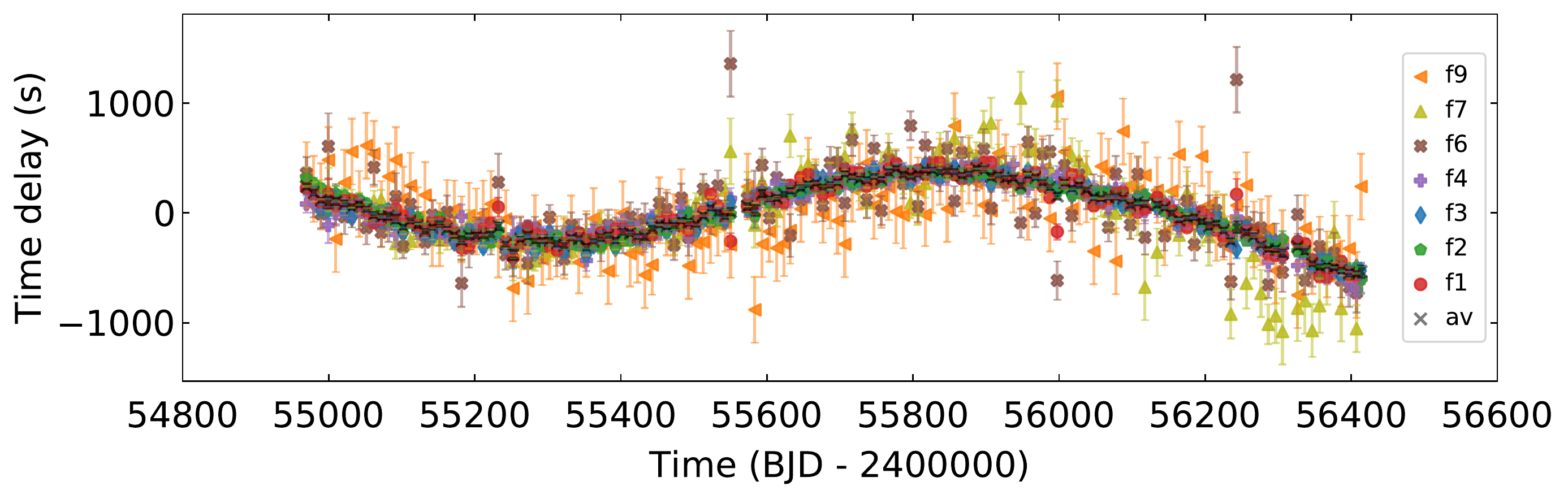}
\includegraphics[width=0.48\textwidth]{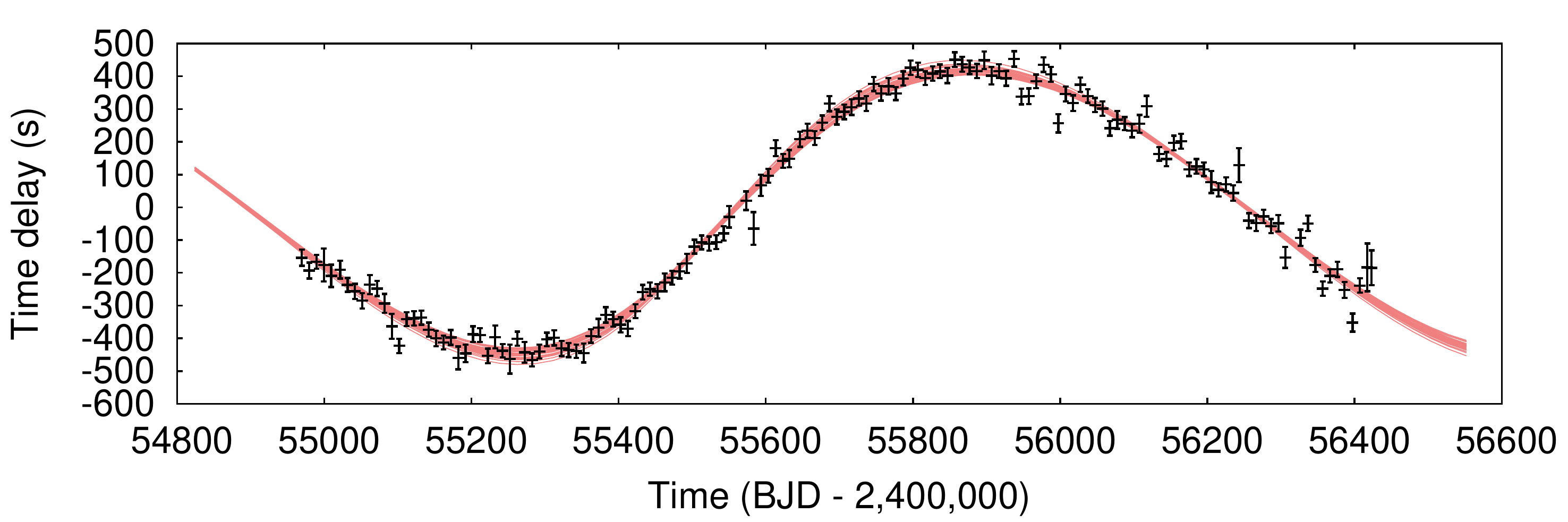}\\
\caption{Time delays for different pulsation modes in five binary stars (left) and orbital solutions for the weighted-average time delay (right), with 25 random steps from the MCMC chain plotted. Targets are (top to bottom): KIC\,11340713, KIC\,3969803, KIC\,4756171, KIC\,5305553, and KIC\,5480040.}
\label{fig:TD_plots}
\end{center}
\end{figure*}

\begin{figure*}
\begin{center}
\includegraphics[width=0.48\textwidth]{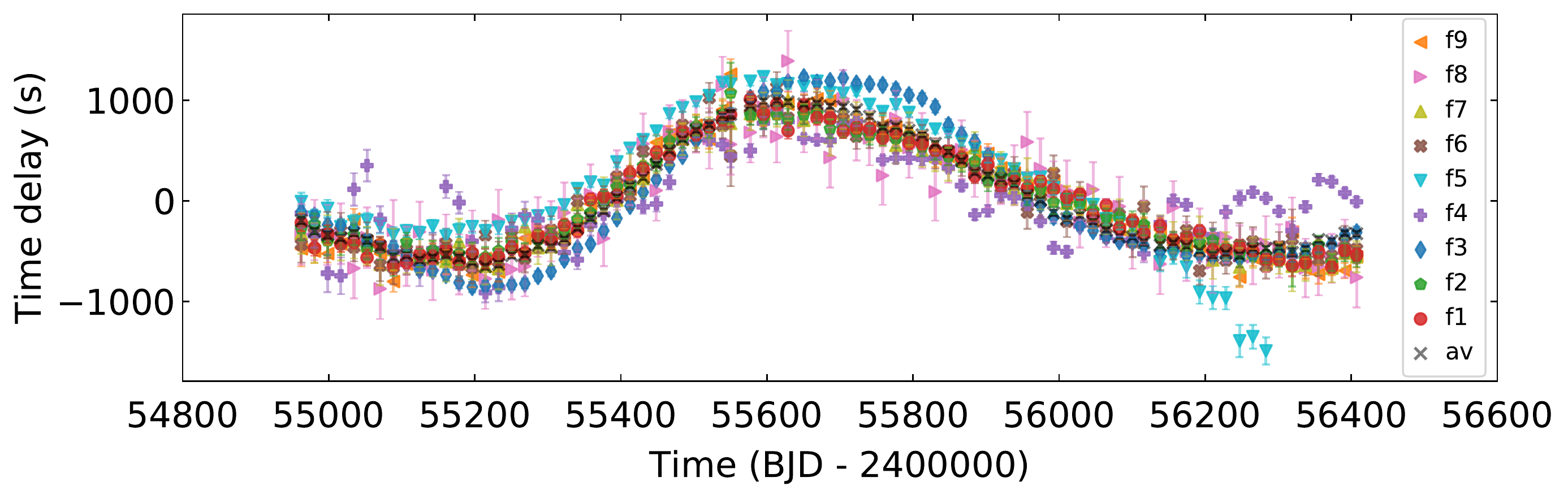}
\includegraphics[width=0.48\textwidth]{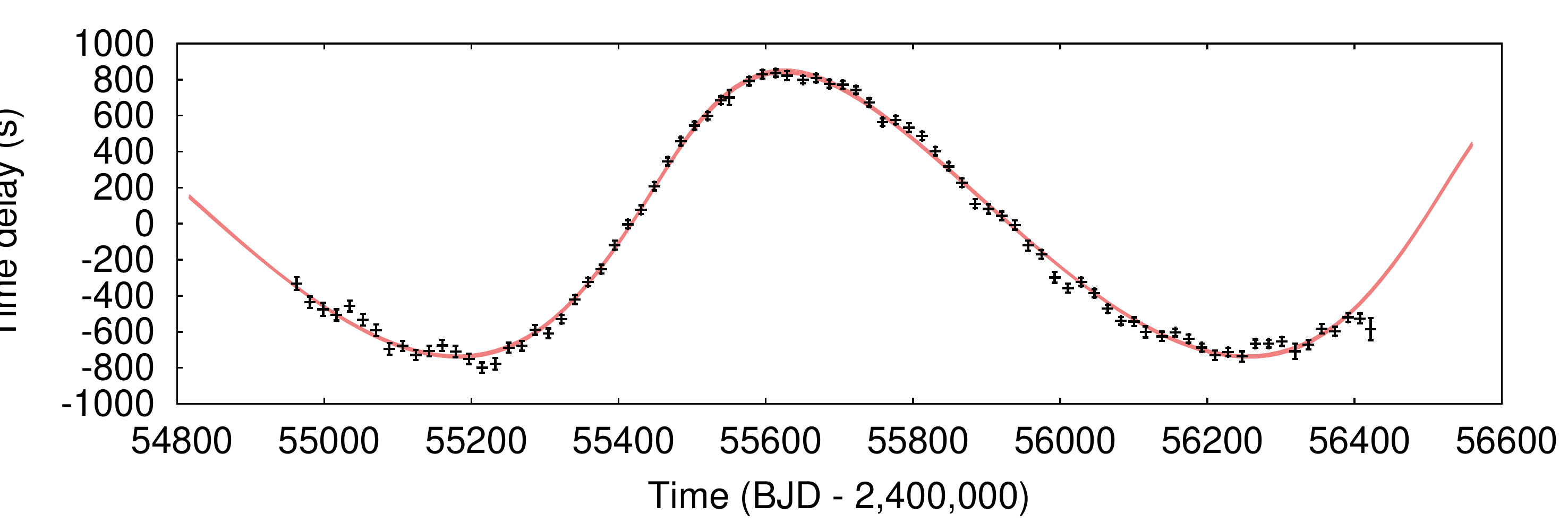}\\
\includegraphics[width=0.48\textwidth]{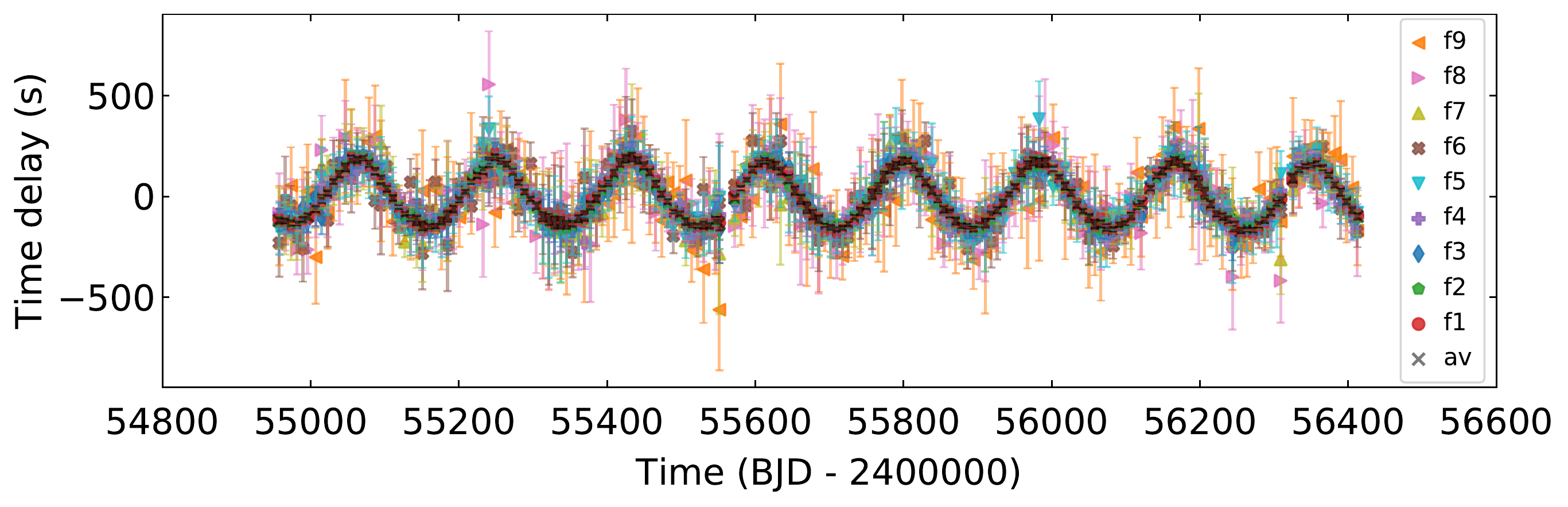}
\includegraphics[width=0.48\textwidth]{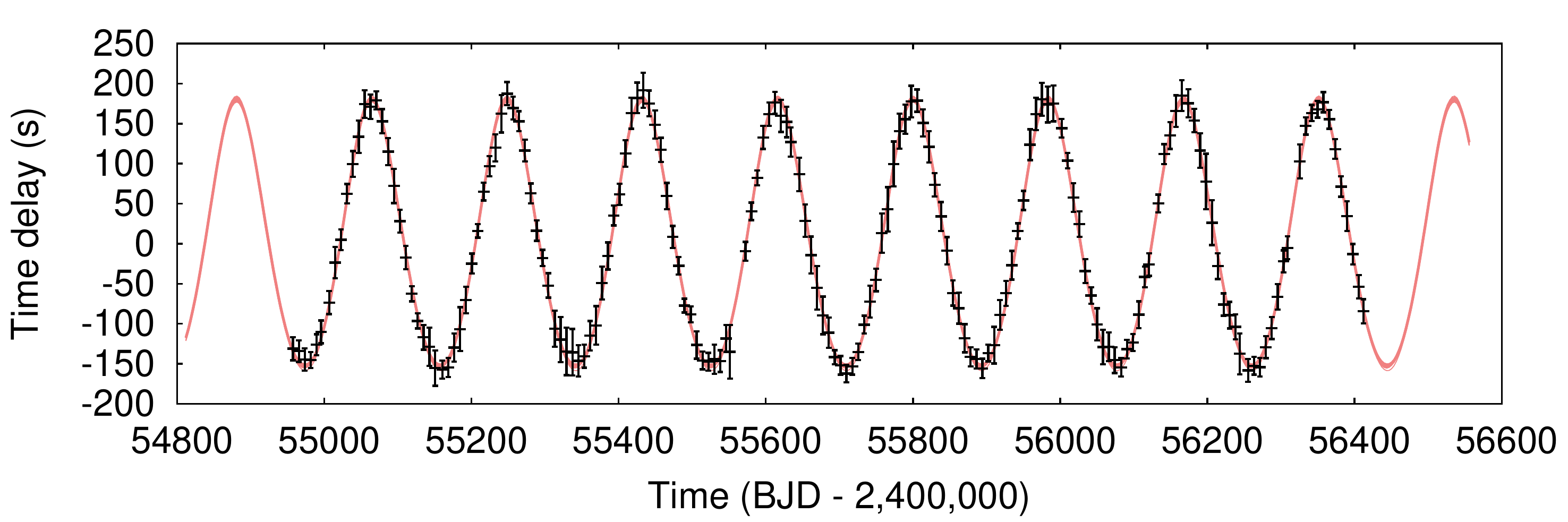}\\
\includegraphics[width=0.48\textwidth]{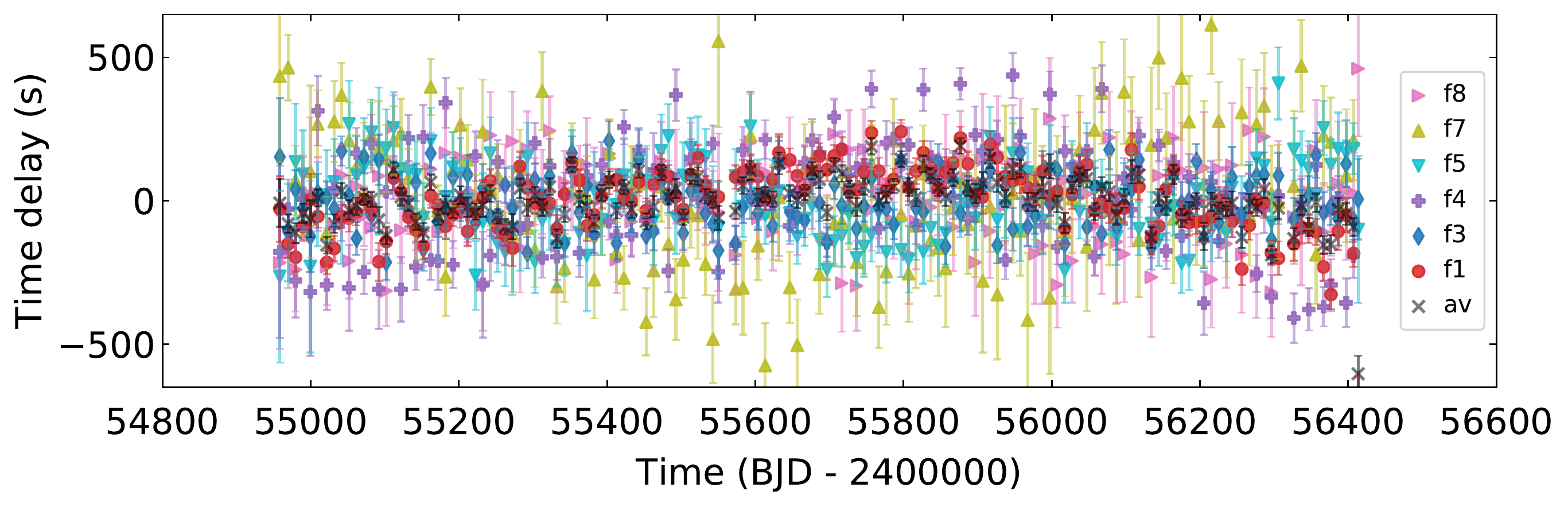}
\includegraphics[width=0.48\textwidth]{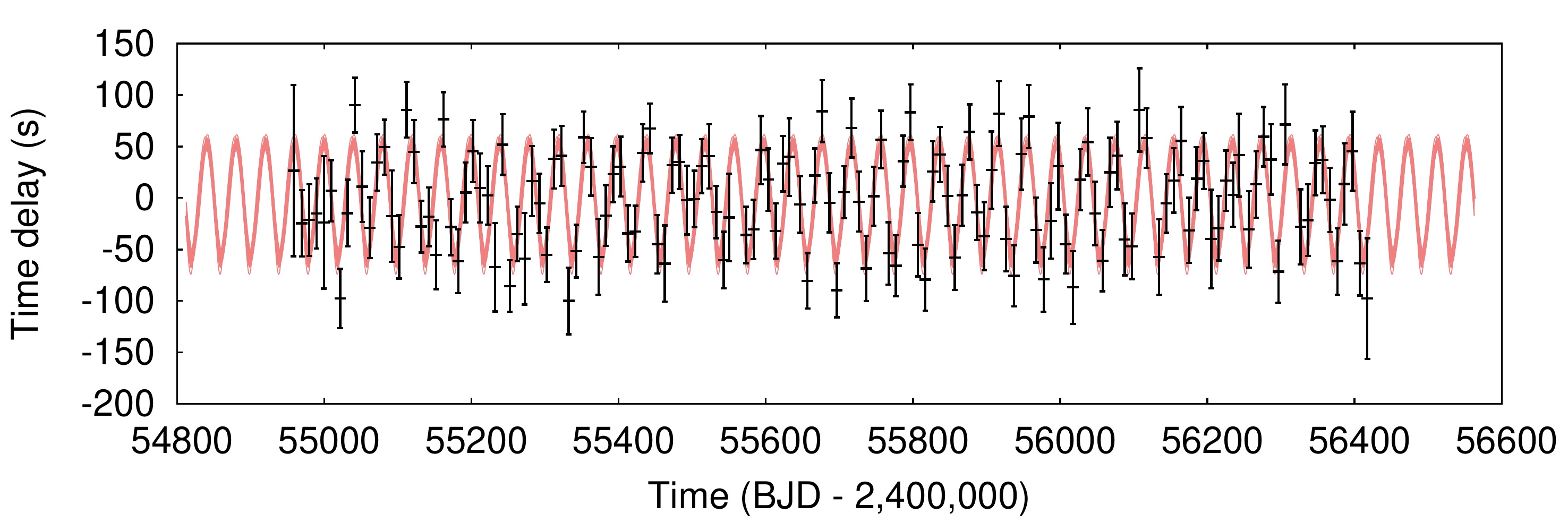}\\
\includegraphics[width=0.48\textwidth]{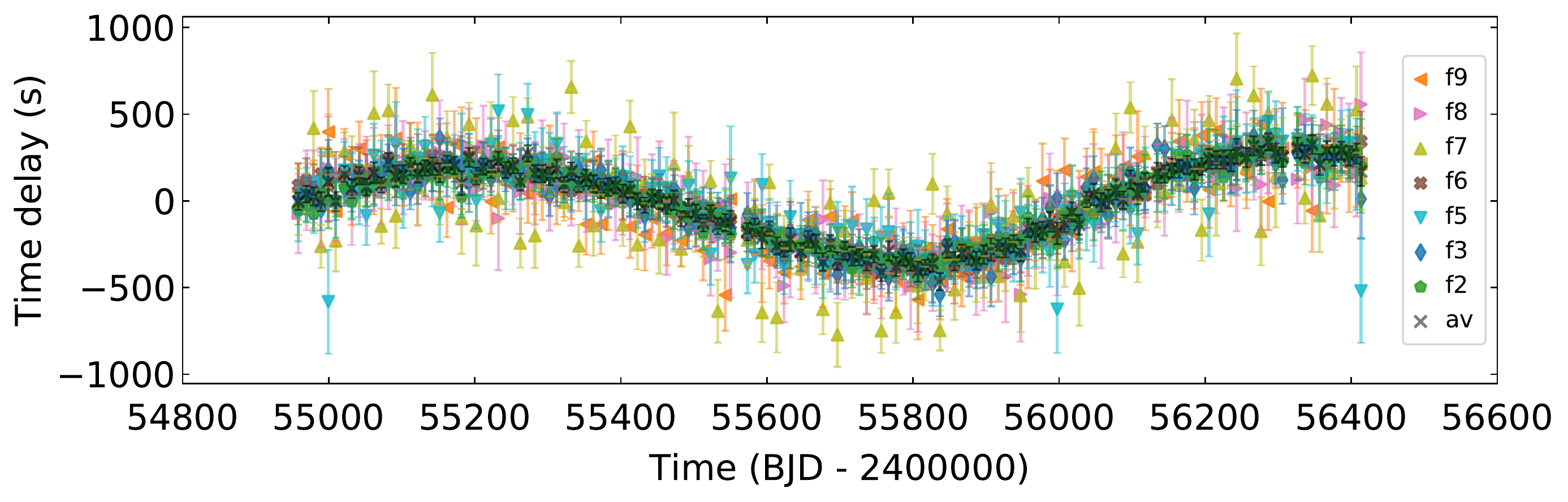}
\includegraphics[width=0.48\textwidth]{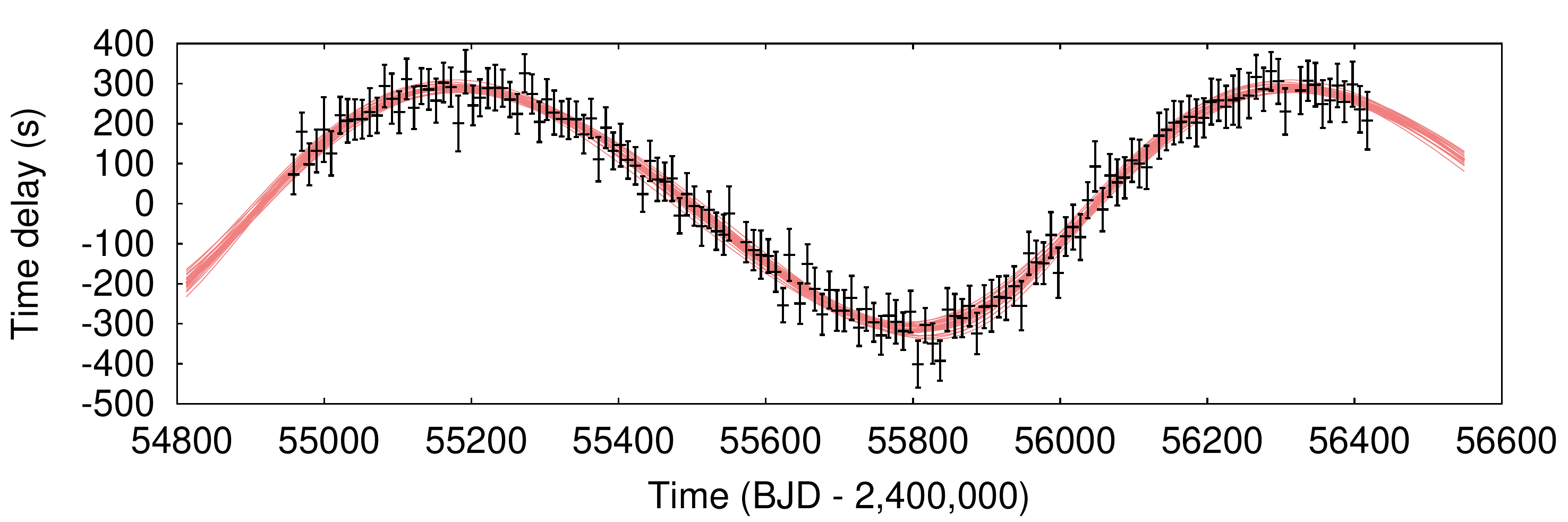}\\
\includegraphics[width=0.48\textwidth]{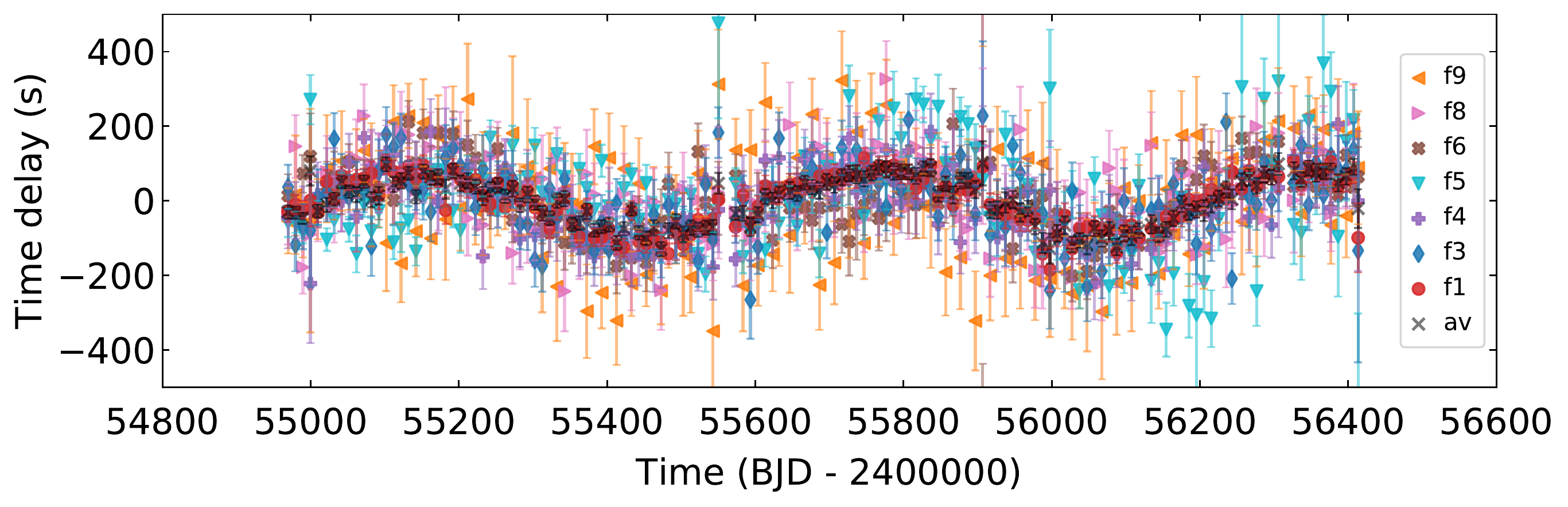}
\includegraphics[width=0.48\textwidth]{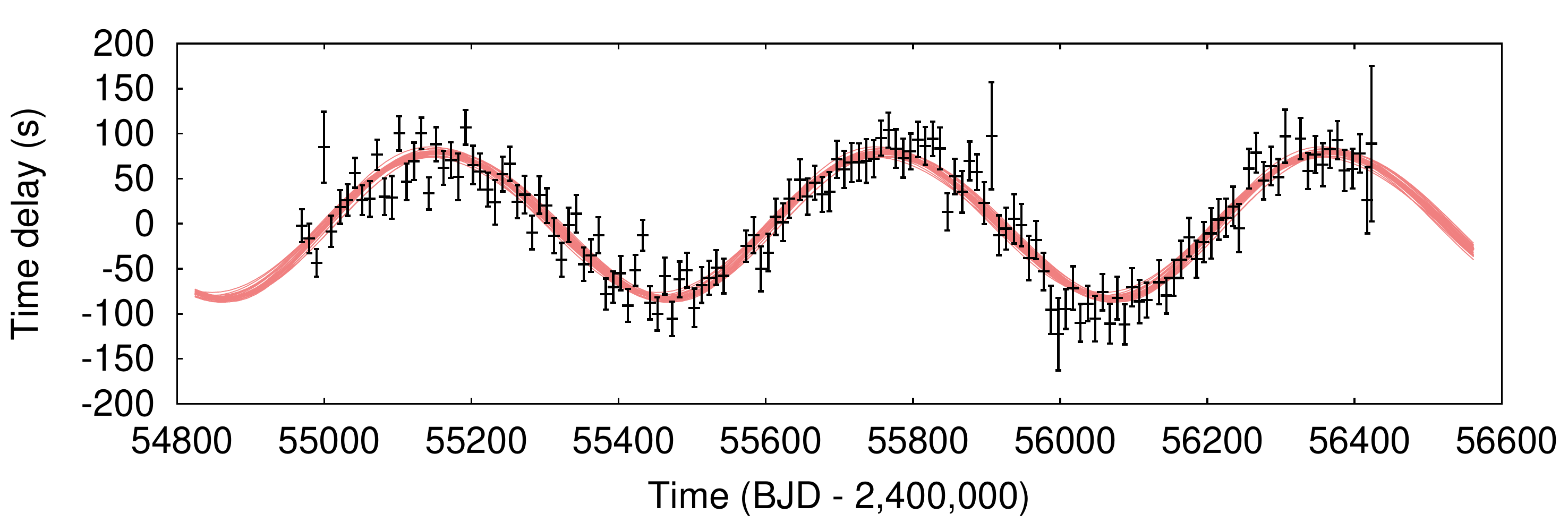}\\
\caption{As Fig.\,\ref{fig:TD_plots}, for five more binaries (top to bottom): KIC\,6804957, KIC\,6887854, KIC\,8647777, KIC\,8842025, and KIC\,9306893.}
\label{fig:TD_plots2}
\end{center}
\end{figure*}

We subdivided the light curves into 10-d subdivisions and, keeping the pulsation frequencies fixed, looked for periodic changes in the pulsation phases that could be attributed to binarity. Each pulsation frequency should generally respond in the same way to the orbit, though minor intrinsic phase modulation occasionally occurs, as in KIC\,11340713 and KIC\,6804957 (Figs\,\ref{fig:TD_plots} and \ref{fig:TD_plots2}). Frequencies with close neighbours in the Fourier transform are strongly affected by beating and were excluded from the analysis (see \citealt{murphyetal2014} for further discussion). After identifying binary candidates, we considered different subdivision sizes that might better sample the orbit, using shorter subdivisions for short-period binaries and longer subdivisions for pulsators with modes that were poorly resolved. The subdivision sizes used for Figs\,\ref{fig:TD_plots} and \ref{fig:TD_plots2} are also given in Table\:\ref{tab:freqs}.

We calculated the weighted-average time delay in each subdivision, weighting by the phase uncertainties for each pulsation mode. These phase uncertainties are the formal least-squares uncertainties for each pulsation mode for each subdivision of the light curve, which also depend on the remaining variance in the light curve (modelled or otherwise). That is, any variance not attributable to the up-to nine fitted pulsation frequencies will lead to larger uncertainties. For this reason, and because lower-frequency (unused) g\:mode pulsations often occur in these stars, we used the {\tt gaussian\_filter} function from {\tt scipy.ndimage} to apply a high-pass gaussian filter to the data to remove this variance. For all but one of the ten discovered binary systems, this led to appropriate uncertainties on the weighted-average time delay, but for KIC\,5305553 the uncertainties remained too large -- almost every time-delay data point was within 1$\sigma$ of the fit (described in the next paragraph). We divided the uncertainties by $4$ for this target, so that the reduced chi-squared ($\chi^2/N$) was close to unity.

To find the best-fitting orbital parameters and their uncertainties, we used a Markov-chain Monte Carlo (MCMC) method \citep{murphyetal2016b} that implements a random walk Metropolis--Hastings algorithm \citep{metropolisetal1953,hastings1970}. The orbits were initialised at values near the expected solution using the semi-analytic method described by \citet{murphy&shibahashi2015}. All orbital solutions were visually inspected to check the quality of the fit and to confirm that $\chi^2/N$ was of order unity.

\begin{table*}
\centering
\caption{Pulsation frequencies extracted from the \textit{Kepler} light curves of the detected binaries. Frequencies excluded from the time-delay averaging are indicated in square brackets. The frequency labels correspond to the ones used in Figs\,\ref{fig:TD_plots} and \ref{fig:TD_plots2}. The final column gives the length of subdivisions used to calculate time delays.}
\label{tab:freqs}
\begin{tabular}{r r r r r r r r r r c }
\toprule
KIC number & $f_1$~~ & $f_2$~~ & $f_3$~~ & $f_4$~~ & $f_5$~~ & $f_6$~~ & $f_7$~~ & $f_8$~~ & $f_9$~~ & subdivision \\
 & d$^{-1}$ & d$^{-1}$ & d$^{-1}$ & d$^{-1}$ & d$^{-1}$ & d$^{-1}$ & d$^{-1}$ & d$^{-1}$ & d$^{-1}$ & d \\
\midrule
11340713 & 10.15 & 9.50 & 10.55 & 9.84 & 11.91 & 13.06 & --- & --- & --- & 10 \\
3969803 & 7.45 & 15.66 & [16.95] & 19.23 & [19.13] & 8.55 & 16.35 & 15.4 & 9.28 & 10 \\
4756171 & 10.68 & 9.98 & 11.79 & 10.38 & 9.66 & [17.49] & 20.41 & 15.86 & [18.25] & \phantom{1}8 \\
5305553 & 16.56 & [32.15] & [5.13] & 19.00 & [5.23] & 17.33 & 17.41 & [5.79] & [4.44] & 10 \\
5480040 & 8.26 & 9.43 & 11.51 & 7.18 & [9.20] & 8.41 & 8.47 & [10.45] & 6.19 & 10 \\
6804957 & 8.24 & 7.15 & 12.72 & [11.06] & 11.66 & 7.22 & 7.92 & 5.19 & 14.11 & 18 \\
6887854 & 28.36 & 29.7 & 36.87 & 27.16 & 29.91 & 35.36 & 36.49 & 36.70 & 29.40 & \phantom{1}8 \\
8647777 & 11.31 & [10.68] & 12.06 & 10.82 & 10.15 & [11.63] & 9.15 & 9.23 & [12.26] & 10 \\
8842025 & [11.47] & 12.72 & 15.13 & [6.25] & 11.11 & 18.92 & 14.63 & 10.37 & 14.17 & 10 \\
9306893 & 16.62 & [29.81] & 16.47 & 15.77 & 31.59 & 25.25 & [17.43] & 19.42 & 19.58 & 10 \\
\bottomrule
\end{tabular}
\end{table*}

\section{Results}
\label{sec:results}

\begin{figure}
\begin{center}
\includegraphics[width=0.48\textwidth]{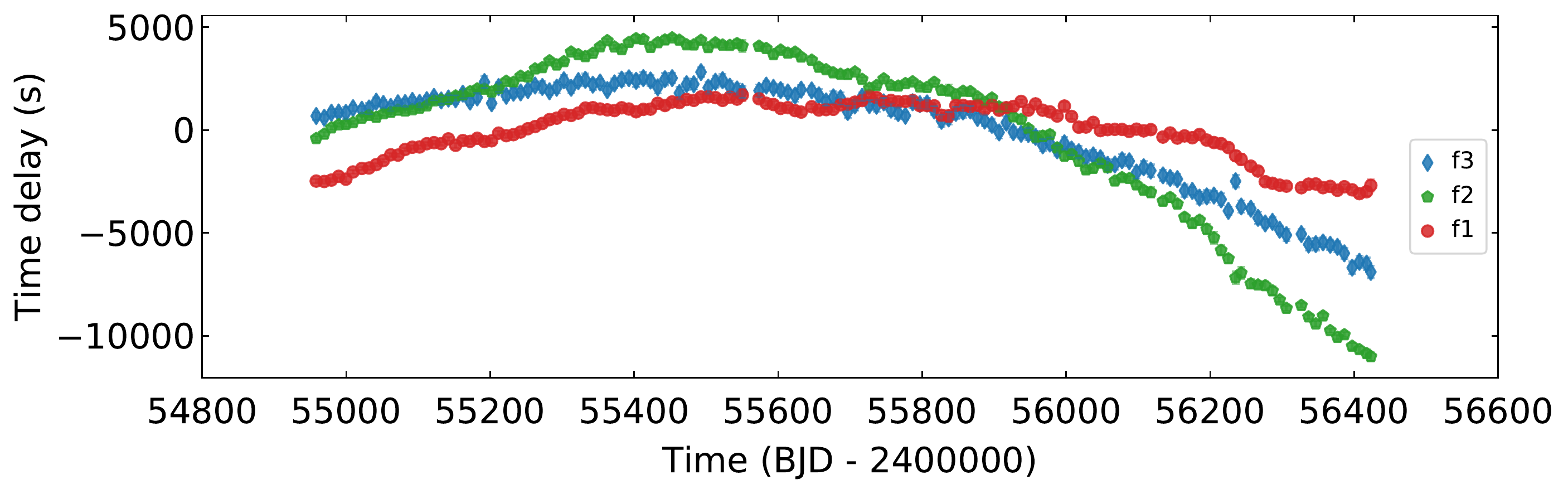}
\caption{The hot star KIC\,7108883, which shows some phase modulation, but the different oscillation modes do not agree well. The pulsation frequencies are $f_1=6.18$, $f_2=8.01$, and $f_3=12.35$\,d$^{-1}$.}
\label{fig:7108883}
\end{center}
\end{figure}

A total of ten new binary systems were discovered (Figures\:\ref{fig:TD_plots} and \ref{fig:TD_plots2}), all of which were from the subsample of 91 cool stars. Thus, 11\% of those targets turned out to be binary systems. Among the hot stars, no new binaries were found, though two targets deserve further mention. Some phase modulation was detected for KIC\,11558725, a known 10-d binary consisting of a white dwarf and a pulsating subdwarf \citep{teltingetal2012}. Our search used only long-cadence data and used 10-d sampling by default, so the 10-d modulation was not detected. Instead, some modulation was detected at longer periods using the g\:modes near 25\,d$^{-1}$, but there was little mutual consistency between different modes. The second target is KIC\,7108883, for which different pulsation modes were also not in good agreement (Fig.\,\ref{fig:7108883}). This target will have to be followed up with radial velocities. If genuine, the binary period is at least as long as the \textit{Kepler} observations.

Given that B stars are known to have a high rate of multiplicity \citep{abtetal1990,sanaetal2013,moe&distefano2017}, it is noteworthy that we find so few in binary systems here. This is presumably due to the fact that B-type stars, like most classes of variable star, are not well-suited to a phase modulation analysis \citep{murphy2018}. Generally, high-frequency oscillators are more useful clocks, but the B stars that do oscillate at high frequencies -- the sdB stars and white dwarfs -- often show intrinsic phase modulation that make them poor clocks \citep{ostensenetal2014a,silvottietal2018,zongetal2018}. If a large sample of $\beta$\,Cep stars with long time-series can be collected with TESS or future missions, these would be the best future B-type targets for phase-modulation studies.

For the ten new binaries, we determine the orbital parameters as the medians of the marginalised posterior distributions from the MCMC chains, which we give in Table\:\ref{tab:orbit}. The uncertainties are given as the upper and lower bounds that encompass the central 68.2\% of the marginalised posteriors.

\begin{table*}
\centering
\caption{Orbital parameters for the PB1 systems. The time of periastron, $t_{\rm p}$, is specified in Barycentric Julian Date. $K_1$ is calculated from the other quantities and is convolved with $\sin i$. The full machine-readable table is available online.}
\label{tab:orbit}
\begin{tabular}{r r@{}l r@{}l r@{}l r@{}l r@{}l r@{}l r@{}l }
\toprule
\multicolumn{1}{c}{KIC number} & \multicolumn{2}{c}{$P$} & \multicolumn{2}{c}{$a_1 \sin i / c$} & \multicolumn{2}{c}{$e$} & \multicolumn{2}{c}{$\varpi$} & \multicolumn{2}{c}{$t_{\rm p}$} & \multicolumn{2}{c}{$f_{\rm M}$} & \multicolumn{2}{c}{$K_1$} \\
\multicolumn{1}{c}{} & \multicolumn{2}{c}{d} & \multicolumn{2}{c}{s} & \multicolumn{2}{c}{} & \multicolumn{2}{c}{rad} & \multicolumn{2}{c}{BJD} & \multicolumn{2}{c}{M$_{\odot}$} & \multicolumn{2}{c}{km\,s$^{-1}$} \\
\midrule
\vspace{1.5mm}
11340713 & $984.8$&$^{+5.0}_{-4.3}$ & $107.35$&$^{+0.98}_{-0.98}$ & $0.102$&$^{+0.018}_{-0.019}$ & $6.100$&$^{+0.058}_{-0.042}$ & $2\,455\,248.0$&$^{+10.0}_{-9.3}$ & $0.00137$&$^{+0.00004}_{-0.00004}$ & $2.389$&$^{+0.027}_{-0.025}$ \\
\vspace{1.5mm}
3969803 & $737$&$^{+18}_{-20}$ & $95.6$&$^{+8.6}_{-7.8}$ & $0.101$&$^{+0.120}_{-0.068}$ & $4.66$&$^{+0.63}_{-0.51}$ & $2\,455\,661$&$^{+80}_{-71}$ & $0.00174$&$^{+0.00051}_{-0.00039}$ & $2.87$&$^{+0.27}_{-0.24}$ \\
\vspace{1.5mm}
4756171 & $31.123$&$^{+0.033}_{-0.034}$ & $66.3$&$^{+8.3}_{-6.6}$ & $0.27$&$^{+0.25}_{-0.18}$ & $4.22$&$^{+0.98}_{-0.62}$ & $2\,454\,984.0$&$^{+4.6}_{-3.4}$ & $0.324$&$^{+0.140}_{-0.087}$ & $48.5$&$^{+11.0}_{-5.5}$ \\
\vspace{1.5mm}
5305553 & $186.73$&$^{+0.20}_{-0.22}$ & $78.7$&$^{+1.4}_{-1.3}$ & $0.200$&$^{+0.034}_{-0.030}$ & $3.215$&$^{+0.076}_{-0.120}$ & $2\,455\,042.2$&$^{+2.6}_{-3.7}$ & $0.01499$&$^{+0.00080}_{-0.00073}$ & $9.37$&$^{+0.21}_{-0.18}$ \\
\vspace{1.5mm}
5480040 & $1358$&$^{+46}_{-34}$ & $440$&$^{+18}_{-14}$ & $0.170$&$^{+0.020}_{-0.026}$ & $2.771$&$^{+0.025}_{-0.049}$ & $2\,455\,484$&$^{+43}_{-43}$ & $0.0493$&$^{+0.0030}_{-0.0023}$ & $7.161$&$^{+0.097}_{-0.088}$ \\
\vspace{1.5mm}
6804957 & $1073.2$&$^{+3.2}_{-3.3}$ & $826.6$&$^{+5.1}_{-5.0}$ & $0.3128$&$^{+0.0097}_{-0.0100}$ & $3.574$&$^{+0.035}_{-0.038}$ & $2\,455\,479.7$&$^{+6.3}_{-5.9}$ & $0.527$&$^{+0.011}_{-0.010}$ & $17.68$&$^{+0.17}_{-0.16}$ \\
\vspace{1.5mm}
6887854 & $183.82$&$^{+0.11}_{-0.12}$ & $167.0$&$^{+1.6}_{-1.7}$ & $0.170$&$^{+0.019}_{-0.018}$ & $4.595$&$^{+0.070}_{-0.071}$ & $2\,455\,061.6$&$^{+2.1}_{-2.2}$ & $0.1481$&$^{+0.0042}_{-0.0045}$ & $20.11$&$^{+0.22}_{-0.24}$ \\
\vspace{1.5mm}
8647777 & $39.813$&$^{+0.041}_{-0.044}$ & $63.9$&$^{+5.3}_{-4.6}$ & $0.20$&$^{+0.18}_{-0.12}$ & $0.57$&$^{+1.30}_{-0.65}$ & $2\,454\,973.4$&$^{+7.4}_{-4.4}$ & $0.177$&$^{+0.047}_{-0.035}$ & $35.8$&$^{+4.2}_{-2.8}$ \\
\vspace{1.5mm}
8842025 & $1133$&$^{+15}_{-13}$ & $307.7$&$^{+7.4}_{-7.5}$ & $0.201$&$^{+0.042}_{-0.043}$ & $2.71$&$^{+0.16}_{-0.11}$ & $2\,455\,977$&$^{+37}_{-29}$ & $0.0243$&$^{+0.0019}_{-0.0018}$ & $6.04$&$^{+0.19}_{-0.19}$ \\
\vspace{1.5mm}
9306893 & $605.9$&$^{+4.3}_{-4.3}$ & $82.0$&$^{+2.7}_{-2.6}$ & $0.079$&$^{+0.055}_{-0.046}$ & $2.41$&$^{+0.31}_{-0.28}$ & $2\,455\,543$&$^{+30}_{-28}$ & $0.00161$&$^{+0.00016}_{-0.00015}$ & $2.96$&$^{+0.10}_{-0.10}$ \\
\bottomrule
\end{tabular}
\end{table*}

All ten of the systems were detected as single-pulsator binaries (PB1s), meaning that the pulsation frequencies arise in the same star. One system (KIC\,8647777) is a possible triple. The orbit in Table\:\ref{tab:orbit} is the short-period one, but there is also a long period variation in the time delays (Fig.\,\ref{fig:TD_plots2}, left). Even the full four-year \textit{Kepler} light curve is not long enough to determine the parameters of the longer orbit. The parameters for the short-period orbit were determined after fitting and subtracting a sinusoid to the long-period variation (Fig.\,\ref{fig:TD_plots2}, right).

\begin{figure}
\begin{center}
\includegraphics[width=0.48\textwidth]{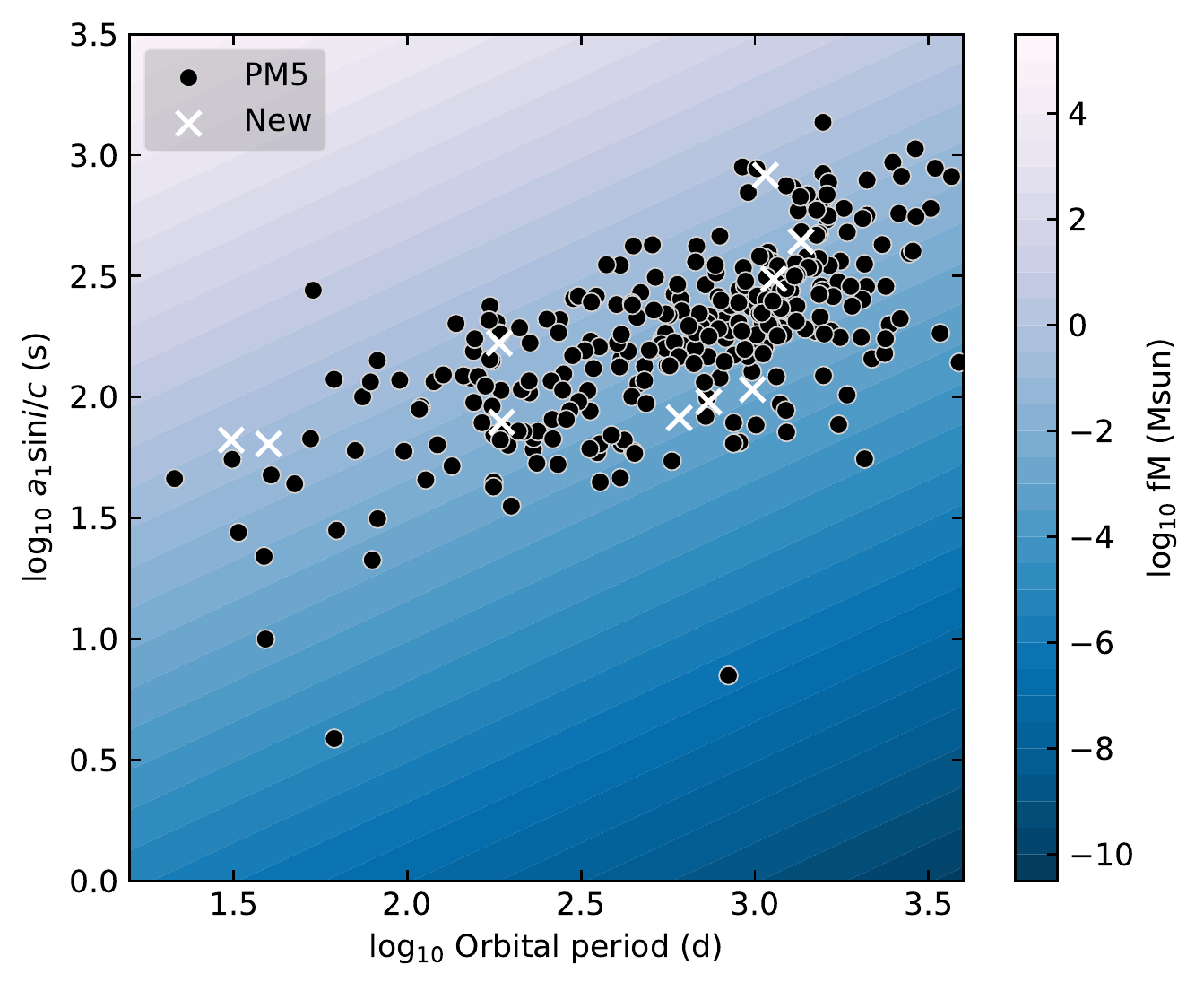}
\caption{Orbital periods and semi-major axes ($a_1 \sin i$/c) of the new systems (cyan crosses) and the binaries from \href{https://ui.adsabs.harvard.edu/abs/2018MNRAS.474.4322M/abstract}{Paper\,V} (black circles). The background shading shows the binary mass function. Adapted from figure 1 of Hey et al. (in press).}
\label{fig:mass_fn}
\end{center}
\end{figure}

Figure\,\ref{fig:mass_fn} shows how the ten new systems fit within the full sample of \textit{Kepler} pulsating binaries. Four of the new systems have relatively large binary mass functions ($f_{\rm M} > 0.1$) and three have relatively small mass functions ($f_{\rm M} \sim 10^{-3}$). All are consistent with being stellar-mass companions, as opposed to brown dwarf or planetary companions.

\begin{figure}
\begin{center}
\includegraphics[width=0.48\textwidth]{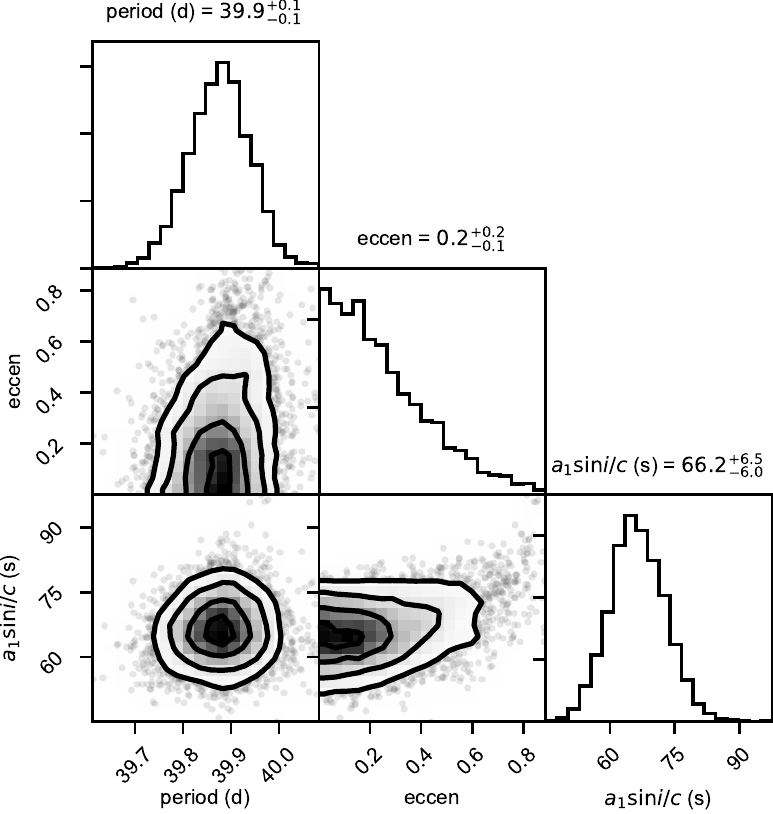}
\caption{Corner plot for the posterior distribution of parameters for KIC\,8647777 using {\sc Maelstrom}. The eccentricity is not well constrained in this short-period system.}
\label{fig:maelstrom}
\end{center}
\end{figure}

The new binaries include two at rather short periods (31 and 40\,d), where the search for binaries with the phase-modulation method has so far proven incomplete (\href{https://ui.adsabs.harvard.edu/abs/2018MNRAS.474.4322M/abstract}{Paper\,V}). It is noteworthy that these two binaries (KIC\,4756161 and 8647777, respectively) also have relatively large mass functions, which made their detection easier. The short-period binaries in \href{https://ui.adsabs.harvard.edu/abs/2018MNRAS.474.4322M/abstract}{Paper\,V} include some systems with low mass functions, so there is no clear implication that the mass ratio distribution at these shorter periods ($P \lesssim 100$\,d) is necessarily different from the distribution at longer periods ($P \gtrsim 100$\,d). To check the accuracy of these two solutions, we also used the {\sc Maelstrom} code (Hey et al., in press) to determine orbits by forward-modelling the time delays onto the light curve. This approach is more sensitive for shorter period and eccentric systems. A comparison of results is given in Table~\ref{tab:maelstrom}. We find that the values agree within the uncertainties, however the eccentricities are still poorly constrained, as indicated by the posterior distribution for one of the stars in Fig.\,\ref{fig:maelstrom}. Radial velocities would help to constrain the eccentricities of these systems.

\begin{table*}
\centering
\caption{Orbital parameter comparison for the short period ($P<100$\,d) binaries found in this work, using the phase modulation method \citep{murphyetal2014,murphyetal2016b} and Maelstrom (Hey et al., submitted).}
\label{tab:maelstrom}
\begin{tabular}{r r@{}l r@{}l r@{}l r@{}l r@{}l r@{}l }
\toprule
\multicolumn{1}{c}{KIC number} & \multicolumn{4}{c}{$P$} & \multicolumn{4}{c}{$a_1 \sin i / c$} & \multicolumn{4}{c}{$e$} \\
\multicolumn{1}{c}{} & \multicolumn{4}{c}{d} & \multicolumn{4}{c}{s} & \multicolumn{4}{c}{} \\
\multicolumn{1}{c}{} & \multicolumn{2}{c}{Phase mod.} & \multicolumn{2}{c}{Maelstrom} & \multicolumn{2}{c}{Phase mod.} & \multicolumn{2}{c}{Maelstrom} & \multicolumn{2}{c}{Phase mod.} & \multicolumn{2}{c}{Maelstrom} \\
\midrule
\vspace{1.5mm}
4756171 & $31.123$&$^{+0.033}_{-0.034}$ & $31.097$&$^{+0.034}_{-0.033}$ & \phantom{00}$66.3$&$^{+8.3}_{-6.6}$ & $63.047$&$^{+5.487}_{-5.258}$ & \phantom{00}$0.27$&$^{+0.25}_{-0.18}$ & $0.158$&$^{+0.165}_{-0.113}$\\
\vspace{1.5mm}
8647777 & $39.813$&$^{+0.041}_{-0.044}$ & $39.878$&$^{+0.062}_{-0.066}$ & \phantom{00}$63.9$&$^{+5.3}_{-4.6}$ & $66.161$&$^{+6.579}_{-6.020}$ & \phantom{00}$0.20$&$^{+0.18}_{-0.12}$ & $0.208$&$^{+0.240}_{-0.146}$ \\
\bottomrule
\end{tabular}
\end{table*}

\begin{figure}
\begin{center}
\includegraphics[width=0.48\textwidth]{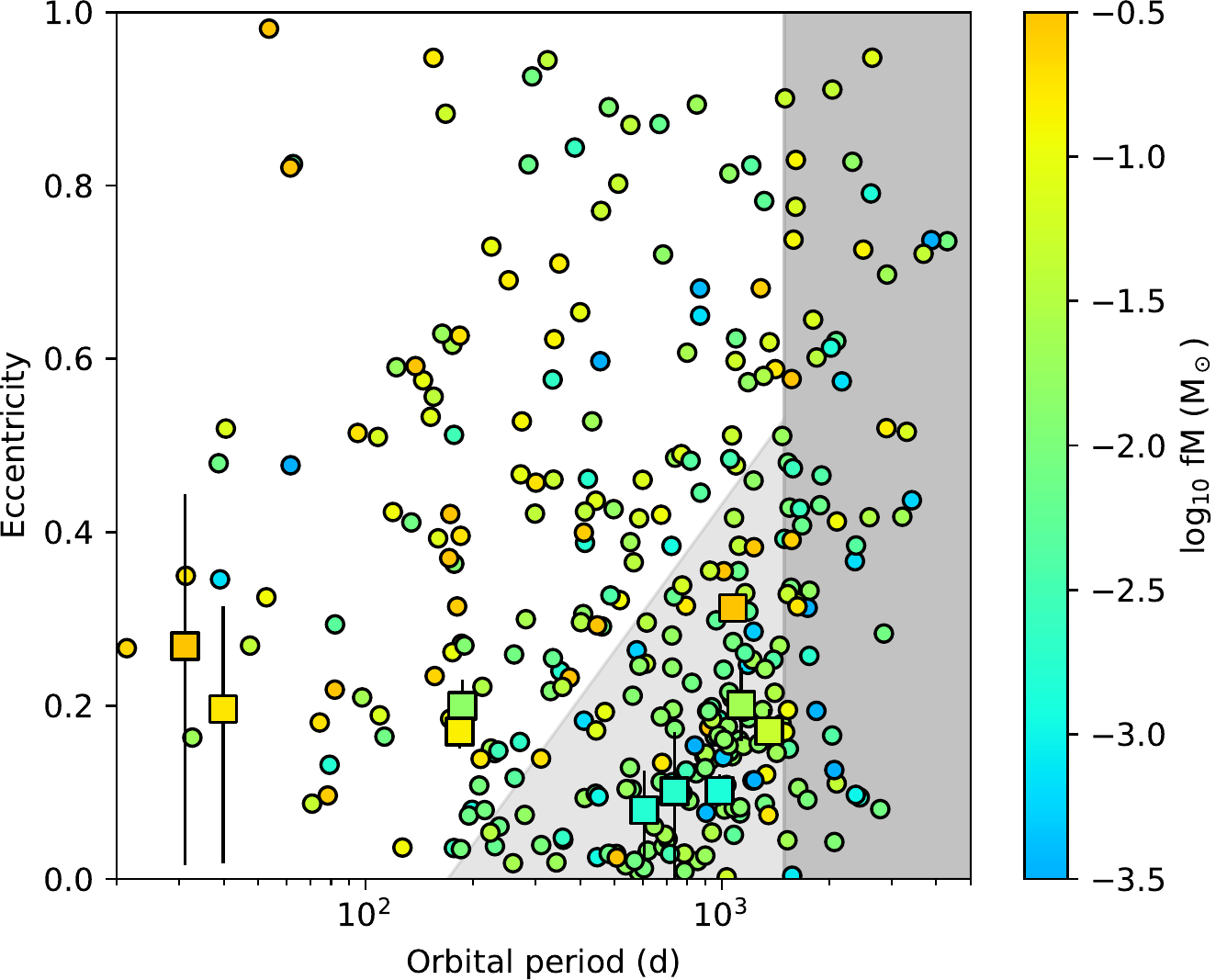}
\caption{The ten new binaries (large squares) compared with the population of binaries from \href{https://ui.adsabs.harvard.edu/abs/2018MNRAS.474.4322M/abstract}{Paper\,V} (small circles) in the period--eccentricity diagram. Systems are colour-coded by their binary mass function, fM. The light-grey triangle is the `post-mass-transfer triangle' defined in figure 9 of \href{https://ui.adsabs.harvard.edu/abs/2018MNRAS.474.4322M/abstract}{Paper\,V}, and the dark grey region contains the system whose orbits cannot be uniquely determined because the period is longer than the \textit{Kepler} data set.}
\label{fig:p-e}
\end{center}
\end{figure}

The largest eccentricity of the ten new binaries is 0.31, and Fig.\,\ref{fig:p-e} shows that six of the ten are located in the long-period, low-eccentricity `post-mass-transfer triangle' (\href{https://ui.adsabs.harvard.edu/abs/2018MNRAS.474.4322M/abstract}{Paper\,V}), where it is possible that the systems have experienced mass transfer. In systems where the original primary has already ascended the asymptotic giant branch (AGB), the AGB star can transfer mass by stable Roche-lobe overflow, causing the binary to circularise and the period to shorten \citep{jorissenetal1998,karakasetal2000,vanwinckel2003,geller&mathieu2011}. Any post-AGB system originally located above or left of this triangle would rapidly become a common-envelope binary instead. Hence systems above or left of this triangle are almost certainly main-sequence pairs, but systems inside the triangle may be post-mass-transfer binaries.
 If these systems have indeed experienced mass transfer, then the system currently consists of a $\delta$\,Sct primary with a white-dwarf secondary whose mass should be around 0.5\,M$_{\odot}$ \citep{jorissenetal1998,vanwinckel2003,geller&mathieu2011}. Such a system would be shaded green in Fig.\,\ref{fig:p-e}.

With a reasonable estimate of the masses of the $\delta$\,Sct primaries, we can estimate the nature of their companions. Given that the systems have been photometrically (mis-)classified with temperatures just cooler than the $\delta$\,Sct red edge, we assume that the primaries are low-mass $\delta$\,Sct stars near the instability strip's red edge, which suggests that they have masses of $\sim$1.5\,M$_{\odot}$ \citep{dupretetal2004,murphyetal2019}. Alternatively, these $\delta$\,Sct stars could have low metallicities and lower masses ($\sim$1.2\,M$_{\odot}$), but that does not greatly affect the following calculations.

We first consider the three systems with very low mass functions (KIC\,11340713, KIC\,3969803 and KIC\,9306893). Only for orbital inclinations below $\sim20^{\circ}$ would the current secondaries have masses consistent with ordinary white dwarfs. At statistically more likely inclinations near $60^{\circ}$, they would have masses closer to 0.2\,M$_{\odot}$, consistent with being extremely low mass white dwarfs (ELMs). However, the formation model for ELMs predicts that they must have orbital periods in the range 2--20\,h \citep{sun&arras2018}, as supported by observations \citep{boffin2015,brownetal2016}. Nonetheless, it is noteworthy that ELMs have been found at longer periods \citep{masudaetal2019}, despite there being no valid formation model for them yet. Determination of the true nature of these three new systems awaits astrometric data from Gaia DR3 to infer the orbital inclinations and hence component masses. They could also be low-mass M dwarfs; distinguishing between these two scenarios observationally is almost impossible.

Of the three other systems in the post-mass-transfer triangle, KIC\,6804957 has a large mass function so it is probably a main-sequence pair occupying this same parameter space; KIC\,8842025 has a mass function consistent with a white dwarf companion; and KIC\,5480040 could have a massive white dwarf companion or be a main-sequence pair. These could be confirmed with high-signal-to-noise spectra, in which case main-sequence companions would be found as double-lined (SB2) systems and white-dwarf companions would be single-lined (SB1).

\section{Conclusions}

We used a machine-learning algorithm to find $\delta$\,Sct stars with mischaracterised temperatures, manually verifying the machine-automated short list of 1203 candidates to recover 960 known and 93 new $\delta$\,Sct stars. We searched these systems for binarity using the phase modulation method and found 10 new binaries. We also searched for binaries among \textit{Kepler} B stars with the same method, but no new binaries could be confirmed. Since B stars are usually found in binary or multiple systems, the low detection rate is probably because B-type variable stars are generally not well-suited to phase modulation analysis.

The new binaries range in orbital period from 31 to 1073\,d. One of the systems may be triple, with an outer period longer than the \textit{Kepler} data-set. The binary mass functions and positions in the period--eccentricity diagram suggest the companions to these $\delta$\,Sct stars are diverse in nature, spanning a wide range of main-sequence masses and possibly also including white dwarfs.

Stars with spectral types between late A and early F are typically rapid rotators, making radial velocity follow-up difficult. Nevertheless, such studies would shed more light on the nature of the companions if the spectra turn out to be double-lined. Otherwise, the orbital parameters correspond (in most cases) to small radial-velocity semi-amplitudes which, along with the typically rapid rotation, suggests that these binaries would not be detectable by conventional methods.

The temperature range that was searched extended down to 6000\,K, considerably cooler than the red-edge of the $\delta$\,Sct instability strip. It is unlikely that $\delta$\,Sct stars with main-sequence companions would have effective temperatures much more mischaracterised than this. However, some $\delta$\,Sct stars have red giant companions that dominate the flux, and such systems would contain a wealth of asteroseismic information. These are promising future targets on which to apply the phase modulation method.

Our application of automated variable-star classification suggests that there is great potential for machine learning in asteroseismology, particularly with wide-field surveys like TESS and PLATO where the number of observed asteroseismic targets is increasing by orders of magnitude.
While manual classification by human experts can be subjective and can contain implicit bias, automated classification methods are objective and can encourage clear, interpretable definitions of the characteristics of a given asteroseismic target.
Although classification algorithms for this application are yet to reach the accuracy of a human expert, even relatively simple algorithms can drastically reduce the manual input required and guide human experts.
This is demonstrated here, where our classification pipeline generated a short list of candidate targets for expert inspection, dramatically reducing the time burden of scouring the full catalogue manually.
As pattern detection algorithms for asteroseismology continue to mature, their ability to guide human expertise is likely to become increasingly valuable in generating new scientific understanding from large and growing asteroseismic datasets.

\section*{Acknowledgements}

This research was supported by the Australian Research Council (ARC). SJM is an ARC DECRA fellow, grant number DE180101104. Funding for the Stellar Astrophysics Centre is provided by the Danish National Research Foundation (grant agreement no.: DNRF106). We thank the anonymous reviewer for suggestions that improved the paper, and thank Kosmas Gazeas for useful comments during the KASC review process.



\bibliographystyle{mnras}
\bibliography{sjm_bibliography} 

\begin{thebibliography}{}
\makeatletter
\relax
\def\mn@urlcharsother{\let\do\@makeother \do\$\do\&\do\#\do\^\do\_\do\%\do\~}
\def\mn@doi{\begingroup\mn@urlcharsother \@ifnextchar [ {\mn@doi@}
  {\mn@doi@[]}}
\def\mn@doi@[#1]#2{\def\@tempa{#1}\ifx\@tempa\@empty \href
  {http://dx.doi.org/#2} {doi:#2}\else \href {http://dx.doi.org/#2} {#1}\fi
  \endgroup}
\def\mn@eprint#1#2{\mn@eprint@#1:#2::\@nil}
\def\mn@eprint@arXiv#1{\href {http://arxiv.org/abs/#1} {{\tt arXiv:#1}}}
\def\mn@eprint@dblp#1{\href {http://dblp.uni-trier.de/rec/bibtex/#1.xml}
  {dblp:#1}}
\def\mn@eprint@#1:#2:#3:#4\@nil{\def\@tempa {#1}\def\@tempb {#2}\def\@tempc
  {#3}\ifx \@tempc \@empty \let \@tempc \@tempb \let \@tempb \@tempa \fi \ifx
  \@tempb \@empty \def\@tempb {arXiv}\fi \@ifundefined
  {mn@eprint@\@tempb}{\@tempb:\@tempc}{\expandafter \expandafter \csname
  mn@eprint@\@tempb\endcsname \expandafter{\@tempc}}}

\bibitem[\protect\citeauthoryear{{Abt}, {Gomez}  \& {Levy}}{{Abt}
  et~al.}{1990}]{abtetal1990}
{Abt} H.~A.,  {Gomez} A.~E.,   {Levy} S.~G.,  1990, \mn@doi [\apjs]
  {10.1086/191508}, \href {http://adsabs.harvard.edu/abs/1990ApJS...74..551A}
  {74, 551}

\bibitem[\protect\citeauthoryear{{Berger}, {Huber}, {Gaidos}  \& {van
  Saders}}{{Berger} et~al.}{2018}]{bergeretal2018}
{Berger} T.~A.,  {Huber} D.,  {Gaidos} E.,   {van Saders} J.~L.,  2018, \mn@doi
  [\apj] {10.3847/1538-4357/aada83}, \href
  {https://ui.adsabs.harvard.edu/abs/2018ApJ...866...99B} {866, 99}

\bibitem[\protect\citeauthoryear{{Boffin}}{{Boffin}}{2015}]{boffin2015}
{Boffin} H.~M.~J.,  2015, \mn@doi [\aap] {10.1051/0004-6361/201525762}, \href
  {http://adsabs.harvard.edu/abs/2015A%26A...575L..13B} {575, L13}

\bibitem[\protect\citeauthoryear{{Borucki} et~al.,}{{Borucki}
  et~al.}{2011}]{boruckietal2011}
{Borucki} W.~J.,  et~al., 2011, \mn@doi [\apj] {10.1088/0004-637X/736/1/19},
  \href {http://adsabs.harvard.edu/abs/2011ApJ...736...19B} {736, 19}

\bibitem[\protect\citeauthoryear{{Brown}, {Latham}, {Everett}  \&
  {Esquerdo}}{{Brown} et~al.}{2011}]{brownetal2011}
{Brown} T.~M.,  {Latham} D.~W.,  {Everett} M.~E.,   {Esquerdo} G.~A.,  2011,
  \mn@doi [\aj] {10.1088/0004-6256/142/4/112}, \href
  {http://adsabs.harvard.edu/abs/2011AJ....142..112B} {142, 112}

\bibitem[\protect\citeauthoryear{{Brown}, {Gianninas}, {Kilic}, {Kenyon}  \&
  {Allende Prieto}}{{Brown} et~al.}{2016}]{brownetal2016}
{Brown} W.~R.,  {Gianninas} A.,  {Kilic} M.,  {Kenyon} S.~J.,   {Allende
  Prieto} C.,  2016, \mn@doi [\apj] {10.3847/0004-637X/818/2/155}, \href
  {http://adsabs.harvard.edu/abs/2016ApJ...818..155B} {818, 155}

\bibitem[\protect\citeauthoryear{{Compton}, {Bedding}, {Murphy}  \&
  {Stello}}{{Compton} et~al.}{2016}]{comptonetal2016}
{Compton} D.~L.,  {Bedding} T.~R.,  {Murphy} S.~J.,   {Stello} D.,  2016,
  \mn@doi [\mnras] {10.1093/mnras/stw1092}, \href
  {http://adsabs.harvard.edu/abs/2016MNRAS.461.1943C} {461, 1943}

\bibitem[\protect\citeauthoryear{{Dupret}, {Grigahc{\`e}ne}, {Garrido},
  {Gabriel}  \& {Scuflaire}}{{Dupret} et~al.}{2004}]{dupretetal2004}
{Dupret} M.-A.,  {Grigahc{\`e}ne} A.,  {Garrido} R.,  {Gabriel} M.,
  {Scuflaire} R.,  2004, \mn@doi [\aap] {10.1051/0004-6361:20031740}, \href
  {http://adsabs.harvard.edu/abs/2004A%26A...414L..17D} {414, L17}

\bibitem[\protect\citeauthoryear{{Dupret}, {Grigahc{\`e}ne}, {Garrido},
  {Gabriel}  \& {Scuflaire}}{{Dupret} et~al.}{2005}]{dupretetal2005b}
{Dupret} M.,  {Grigahc{\`e}ne} A.,  {Garrido} R.,  {Gabriel} M.,   {Scuflaire}
  R.,  2005, \mn@doi [\aap] {10.1051/0004-6361:20041817}, \href
  {http://adsabs.harvard.edu/abs/2005A%26A...435..927D} {435, 927}

\bibitem[\protect\citeauthoryear{{Fulcher} \& {Jones}}{{Fulcher} \&
  {Jones}}{2014}]{fulcher&jones2014}
{Fulcher} B.~D.,  {Jones} N.~S.,  2014, IEEE TRANSACTIONS ON KNOWLEDGE AND DATA
  ENGINEERING, \href {https://ui.adsabs.harvard.edu/abs/2014arXiv1401.3531F}
  {26, 3026}

\bibitem[\protect\citeauthoryear{{Fulcher} \& {Jones}}{{Fulcher} \&
  {Jones}}{2017}]{fulcher&jones2017}
{Fulcher} B.~D.,  {Jones} N.~S.,  2017, \mn@doi [Cell Systems]
  {https://doi.org/10.1016/j.cels.2017.10.001}, 5, 527

\bibitem[\protect\citeauthoryear{{Fulcher}, {Little}  \& {Jones}}{{Fulcher}
  et~al.}{2013}]{fulcheretal2013}
{Fulcher} B.~D.,  {Little} M.~A.,   {Jones} N.~S.,  2013, \mn@doi [Journal of
  the Royal Society] {https://doi.org/10.1098/rsif.2013.0048}, \href
  {https://ui.adsabs.harvard.edu/abs/2013arXiv1304.1209F} {10}

\bibitem[\protect\citeauthoryear{{Geller} \& {Mathieu}}{{Geller} \&
  {Mathieu}}{2011}]{geller&mathieu2011}
{Geller} A.~M.,  {Mathieu} R.~D.,  2011, \mn@doi [\nat] {10.1038/nature10512},
  \href {http://adsabs.harvard.edu/abs/2011Natur.478..356G} {478, 356}

\bibitem[\protect\citeauthoryear{{Gilliland} et~al.,}{{Gilliland}
  et~al.}{2010}]{gillilandetal2010a}
{Gilliland} R.~L.,  et~al., 2010, \mn@doi [\pasp] {10.1086/650399}, \href
  {http://adsabs.harvard.edu/abs/2010PASP..122..131G} {122, 131}

\bibitem[\protect\citeauthoryear{{Hastings}}{{Hastings}}{1970}]{hastings1970}
{Hastings} W.~K.,  1970, Biometrika, \href
  {http://adsabs.harvard.edu/abs/1970CoAsi.239....1B} {57, 97}

\bibitem[\protect\citeauthoryear{{Huber} et~al.,}{{Huber}
  et~al.}{2014}]{huberetal2014}
{Huber} D.,  et~al., 2014, \mn@doi [\apjs] {10.1088/0067-0049/211/1/2}, \href
  {http://adsabs.harvard.edu/abs/2014ApJS..211....2H} {211, 2}

\bibitem[\protect\citeauthoryear{{Jorissen}, {Van Eck}, {Mayor}  \&
  {Udry}}{{Jorissen} et~al.}{1998}]{jorissenetal1998}
{Jorissen} A.,  {Van Eck} S.,  {Mayor} M.,   {Udry} S.,  1998, \aap, \href
  {http://adsabs.harvard.edu/abs/1998A%26A...332..877J} {332, 877}

\bibitem[\protect\citeauthoryear{{Karakas}, {Tout}  \& {Lattanzio}}{{Karakas}
  et~al.}{2000}]{karakasetal2000}
{Karakas} A.~I.,  {Tout} C.~A.,   {Lattanzio} J.~C.,  2000, \mn@doi [\mnras]
  {10.1046/j.1365-8711.2000.03561.x}, \href
  {http://adsabs.harvard.edu/abs/2000MNRAS.316..689K} {316, 689}

\bibitem[\protect\citeauthoryear{{Masuda}, {Kawahara}, {Latham}, {Bieryla},
  {Kunitomo}, {MacLeod}  \& {Aoki}}{{Masuda} et~al.}{2019}]{masudaetal2019}
{Masuda} K.,  {Kawahara} H.,  {Latham} D.~W.,  {Bieryla} A.,  {Kunitomo} M.,
  {MacLeod} M.,   {Aoki} W.,  2019, \mn@doi [\apjl] {10.3847/2041-8213/ab321b},
  \href {https://ui.adsabs.harvard.edu/abs/2019ApJ...881L...3M} {881, L3}

\bibitem[\protect\citeauthoryear{{Mathur} et~al.,}{{Mathur}
  et~al.}{2017}]{mathuretal2017}
{Mathur} S.,  et~al., 2017, \mn@doi [\apjs] {10.3847/1538-4365/229/2/30}, \href
  {http://adsabs.harvard.edu/abs/2017ApJS..229...30M} {229, 30}

\bibitem[\protect\citeauthoryear{{Metropolis}, {Rosenbluth}, {Rosenbluth},
  {Teller}  \& {Teller}}{{Metropolis} et~al.}{1953}]{metropolisetal1953}
{Metropolis} N.,  {Rosenbluth} A.,  {Rosenbluth} M.,  {Teller} A.,   {Teller}
  E.,  1953, J. Chem. Phys., 21, 1087

\bibitem[\protect\citeauthoryear{{Moe} \& {Di Stefano}}{{Moe} \& {Di
  Stefano}}{2017}]{moe&distefano2017}
{Moe} M.,  {Di Stefano} R.,  2017, \mn@doi [\apjs] {10.3847/1538-4365/aa6fb6},
  \href {http://adsabs.harvard.edu/abs/2017ApJS..230...15M} {230, 15}

\bibitem[\protect\citeauthoryear{{Murphy}}{{Murphy}}{2018}]{murphy2018}
{Murphy} S.~J.,  2018, arXiv e-prints, \href
  {https://ui.adsabs.harvard.edu/abs/2018arXiv181112659M} {p. arXiv:1811.12659}

\bibitem[\protect\citeauthoryear{{Murphy} \& {Shibahashi}}{{Murphy} \&
  {Shibahashi}}{2015}]{murphy&shibahashi2015}
{Murphy} S.~J.,  {Shibahashi} H.,  2015, \mn@doi [\mnras]
  {10.1093/mnras/stv884}, \href
  {http://adsabs.harvard.edu/abs/2015MNRAS.450.4475M} {450, 4475}

\bibitem[\protect\citeauthoryear{{Murphy}, {Bedding}, {Shibahashi}, {Kurtz}  \&
  {Kjeldsen}}{{Murphy} et~al.}{2014}]{murphyetal2014}
{Murphy} S.~J.,  {Bedding} T.~R.,  {Shibahashi} H.,  {Kurtz} D.~W.,
  {Kjeldsen} H.,  2014, \mn@doi [\mnras] {10.1093/mnras/stu765}, \href
  {http://adsabs.harvard.edu/abs/2014MNRAS.441.2515M} {441, 2515}

\bibitem[\protect\citeauthoryear{{Murphy}, {Shibahashi}  \& {Bedding}}{{Murphy}
  et~al.}{2016}]{murphyetal2016b}
{Murphy} S.~J.,  {Shibahashi} H.,   {Bedding} T.~R.,  2016, \mn@doi [\mnras]
  {10.1093/mnras/stw1622}, \href
  {http://adsabs.harvard.edu/abs/2016MNRAS.461.4215M} {461, 4215}

\bibitem[\protect\citeauthoryear{{Murphy}, {Moe}, {Kurtz}, {Bedding},
  {Shibahashi}  \& {Boffin}}{{Murphy} et~al.}{2018}]{murphyetal2018}
{Murphy} S.~J.,  {Moe} M.,  {Kurtz} D.~W.,  {Bedding} T.~R.,  {Shibahashi} H.,
   {Boffin} H.~M.~J.,  2018, \mn@doi [\mnras] {10.1093/mnras/stx3049}, \href
  {http://adsabs.harvard.edu/abs/2018MNRAS.474.4322M} {474, 4322}

\bibitem[\protect\citeauthoryear{{Murphy}, {Hey}, {Van Reeth}  \&
  {Bedding}}{{Murphy} et~al.}{2019}]{murphyetal2019}
{Murphy} S.~J.,  {Hey} D.,  {Van Reeth} T.,   {Bedding} T.~R.,  2019, \mn@doi
  [\mnras] {10.1093/mnras/stz590}, \href
  {https://ui.adsabs.harvard.edu/abs/2019MNRAS.485.2380M} {485, 2380}

\bibitem[\protect\citeauthoryear{{{\O}stensen}, {Reed}, {Baran}  \&
  {Telting}}{{{\O}stensen} et~al.}{2014}]{ostensenetal2014a}
{{\O}stensen} R.~H.,  {Reed} M.~D.,  {Baran} A.~S.,   {Telting} J.~H.,  2014,
  \mn@doi [\aap] {10.1051/0004-6361/201423734}, \href
  {http://adsabs.harvard.edu/abs/2014A%26A...564L..14O} {564, L14}

\bibitem[\protect\citeauthoryear{{Sana} et~al.,}{{Sana}
  et~al.}{2013}]{sanaetal2013}
{Sana} H.,  et~al., 2013, \mn@doi [\aap] {10.1051/0004-6361/201219621}, \href
  {http://adsabs.harvard.edu/abs/2013A%26A...550A.107S} {550, A107}

\bibitem[\protect\citeauthoryear{{Shibahashi} \& {Kurtz}}{{Shibahashi} \&
  {Kurtz}}{2012}]{shibahashi&kurtz2012}
{Shibahashi} H.,  {Kurtz} D.~W.,  2012, \mn@doi [\mnras]
  {10.1111/j.1365-2966.2012.20654.x}, \href
  {http://adsabs.harvard.edu/abs/2012MNRAS.422..738S} {422, 738}

\bibitem[\protect\citeauthoryear{{Silvotti} et~al.,}{{Silvotti}
  et~al.}{2018}]{silvottietal2018}
{Silvotti} R.,  et~al., 2018, \mn@doi [\aap] {10.1051/0004-6361/201731473},
  \href {http://adsabs.harvard.edu/abs/2018A%26A...611A..85S} {611, A85}

\bibitem[\protect\citeauthoryear{{Stumpe}, {Smith}, {Catanzarite}, {Van Cleve},
  {Jenkins}, {Twicken}  \& {Girouard}}{{Stumpe} et~al.}{2014}]{stumpeetal2014}
{Stumpe} M.~C.,  {Smith} J.~C.,  {Catanzarite} J.~H.,  {Van Cleve} J.~E.,
  {Jenkins} J.~M.,  {Twicken} J.~D.,   {Girouard} F.~R.,  2014, \mn@doi [\pasp]
  {10.1086/674989}, \href {http://adsabs.harvard.edu/abs/2014PASP..126..100S}
  {126, 100}

\bibitem[\protect\citeauthoryear{{Sun} \& {Arras}}{{Sun} \&
  {Arras}}{2018}]{sun&arras2018}
{Sun} M.,  {Arras} P.,  2018, \mn@doi [\apj] {10.3847/1538-4357/aab9a4}, \href
  {http://adsabs.harvard.edu/abs/2018ApJ...858...14S} {858, 14}

\bibitem[\protect\citeauthoryear{{Telting} et~al.,}{{Telting}
  et~al.}{2012}]{teltingetal2012}
{Telting} J.~H.,  et~al., 2012, \mn@doi [\aap] {10.1051/0004-6361/201219458},
  \href {http://adsabs.harvard.edu/abs/2012A%26A...544A...1T} {544, A1}

\bibitem[\protect\citeauthoryear{{Zong}, {Charpinet}, {Fu}, {Vauclair}, {Niu}
  \& {Su}}{{Zong} et~al.}{2018}]{zongetal2018}
{Zong} W.,  {Charpinet} S.,  {Fu} J.-N.,  {Vauclair} G.,  {Niu} J.-S.,   {Su}
  J.,  2018, \mn@doi [\apj] {10.3847/1538-4357/aaa548}, \href
  {http://adsabs.harvard.edu/abs/2018ApJ...853...98Z} {853, 98}

\bibitem[\protect\citeauthoryear{{van Winckel}}{{van
  Winckel}}{2003}]{vanwinckel2003}
{van Winckel} H.,  2003, \mn@doi [\araa]
  {10.1146/annurev.astro.41.071601.170018}, \href
  {http://adsabs.harvard.edu/abs/2003ARA%26A..41..391V} {41, 391}

\makeatother
\end{thebibliography}



\bsp	
\label{lastpage}
\end{document}